\documentclass{nature}
\usepackage{times}
\usepackage{url}
\usepackage{xcolor}
\usepackage{hyperref}
\usepackage{amsmath}
\usepackage{amssymb}
\usepackage{amsfonts}
\usepackage{booktabs}
\usepackage{graphicx}
\graphicspath{{figs/}}

\usepackage{lineno}
\usepackage{dcolumn} 
\newcolumntype{d}[1]{D{.}{.}{#1}}

\usepackage[table-number-alignment=center, group-separator={}]{siunitx}
\def\sym#1{\ifmmode^{#1}\else\(^{#1}\)\fi}
\usepackage{makecell}

\makeatletter
\let\saved@includegraphics\includegraphics
\AtBeginDocument{\let\includegraphics\saved@includegraphics}
\makeatother

\newcommand{\sm}[1]{\textcolor{blue}{SI #1}}
\newcommand{\fref}[1]{Fig.~\ref{#1}}

\title{\large{The Gender Gap in Scholarly Self-Promotion on Social Media}}

\author{Hao Peng$^{1,2,3}$, Misha Teplitskiy$^{1,4}$, Daniel M. Romero$^{1,5,6}$, Em\H oke-\'Agnes Horv\'at$^{3,7,8}$}

\begin{document}

\maketitle

\begin{affiliations} 
\item School of Information, University of Michigan, Ann Arbor, MI 48109, USA
\item Kellogg School of Management, Northwestern University, Evanston, IL 60208, USA
\item Northwestern University Institute on Complex Systems, Evanston, IL 60208, USA
\item Laboratory for Innovation Science at Harvard, Harvard University, Boston, MA, USA
\item Center for the Study of Complex Systems, University of Michigan, Ann Arbor, MI 48109, USA
\item Department of Electrical Engineering and Computer Science, University of Michigan, MI, USA
\item School of Communication, Northwestern University, Evanston, IL 60208, USA
\item McCormick School of Engineering, Northwestern University, Evanston, IL 60208, USA
\end{affiliations} 

\begin{abstract}
Self-promotion in science is ubiquitous but may not be exercised equally by men and women. Research on self-promotion in other domains suggests that, due to bias in self-assessment and adverse reactions to non-gender-conforming behaviors (``pushback''), women tend to self-promote less often than men. We test whether this pattern extends to scholars by examining self-promotion over six years using 23M Tweets about 2.8M research papers by 3.5M authors. 
Overall, women are about 28\% less likely than men to self-promote their papers even after accounting for important confounds, and this gap has grown over time.
Moreover, differential adoption of Twitter does not explain the gender gap, which is large even in relatively gender-balanced broad research areas, where bias in self-assessment and pushback are expected to be smaller.
Further, the gap increases with higher performance and status, being most pronounced for productive women from top-ranked institutions who publish in high-impact journals.
Critically, we find differential returns with respect to gender: while self-promotion is associated with increased tweets of papers, the increase is smaller for women than for men. 
Our findings suggest that self-promotion varies meaningfully by gender and help explain gender differences in the visibility of scientific ideas.
\end{abstract}
\newpage

\section{Introduction}

Traditional and social media play an important role in the dissemination of scientific findings \cite{ncb2018social, peters2013gap}.
For example, an established line of literature suggests that the online visibility of scholarly papers amplifies their impact in the academy through citations \cite{costas2015altmetrics, dehdarirad2020could, klar2020using, luc2021does} but also beyond academic audiences \cite{eagleman2013public, sugimoto2017scholarly}, contributing to prevalent alternative measures for scholarly evaluation \cite{kwok2013research}.
Social media platforms are thus commonly used by scholars across disciplines to discuss ideas and increase their visibility \cite{darling2013role, morello2015we, gero2021makes}.
Recent research reveals that gender gaps observed in traditional scientific outcomes like citations, hiring, and awards \cite{sugimoto2023equity, way2016gender, oliveira2019comparison} also appear in online visibility \cite{vasarhelyi2021gender}, despite the absence of traditional gate-keepers. What explains this gender gap in online visibility? We consider one possible mechanism---self-promotion. 

Self-promotion refers to speaking directly about one's strengths and achievements in professional contexts and has long been recognized as a career-enhancing practice \cite{schlenker1980impression}. 
In general, individuals who engage in self-promotion do so in order to be perceived as competent, ambitious, and confident, which are required for effective leadership and career progression \cite{stevens1995making, leary2019self}. 
Self-promotion in science may be particularly consequential in the early stage of research dissemination, where small initial differences in visibility can accumulate into substantial ones over time \cite{lieberman1988first, bol2018matthew}. 

However, several literatures suggest that men and women tend to engage in self-promotion at different rates. 
First, in order to promote one's achievements, one must recognize them as worthy of promotion. A number of studies find gender bias in such self-assessments, with women often undervaluing their performance in stereotypically male-dominated domains like science and business, and men overvaluing theirs \cite{coffman2014evidence, josephs1992gender, bleidorn2016age, correll2001gender}.
For example, a recent study by Exley and Kessler shows that women and girls self-assess the same performance on math tasks lower than men and boys, and the gap arises as early as sixth grade \cite{exley2019gender}. 
The second literature concerns the ``backlash effect,'' whereby audiences differentially penalize women for self-promotion because it is incongruent with stereotypes of modesty and pro-sociality  \cite{phelan2010prejudice, rudman2008backlash, moss2010disruptions, daubman1992gender, heatherington1993two, shor2022women}.
Supporting this view on double standards, empirical research across many settings finds that women who self-promote are often seen as more arrogant and less likable than self-promoting men \cite{hagen1975discrimination, rudman2001prescriptive, janoff1996dilemma, rudman1998self, altenburger2017there, brooks2014investors}. 
Additionally, studies of online harassment, even in non-self-promotional contexts, find gender differences in how frequently harassment occurs. 
By one estimate, 85\% of women and girls globally have experienced some form of online harassment and abuse \cite{crockett2022}. 
Harassment is common on social media sites, with 61\% of women considering it a major problem \cite{vogels2021state}, and may naturally discourage posting. Recent research documents how gender harassment has changed how women popularize science \cite{mcdonald2020effects}.
Third, self-promotion takes time. A growing literature documents that women spend more time on non-promotable tasks, such as emotional support towards trainees \cite{babcock2017gender, nelson2023taking} and academic services \cite{miller2020balancing}.
Thus gender differences in time allocated to research may give men more bandwidth to self-promote. 

These three strands of literature suggest that men are likely to self-promote their scientific work more than women, and lead to our first research question: \emph{Is there a gender gap in the self-promotion of science on social media?} Additionally, we explore whether the gap has changed over time, and whether it appears in all areas of science, including those with better gender representation.

Informing interventions to close a potential gender gap in self-promotion requires understanding crucial nuances of who is more likely to engage in self-promotion.
On the one hand, self-assessment biases are likely to be smaller for high-status individuals and/or when the performance is more unambiguously good \cite{exley2019gender}.
On the other hand, past research shows that high-achieving women can elicit more pushback for their success \cite{cooper2013women}, for instance, experiencing less positive sentiment in their media coverage \cite{shor2022women}.
The net effect of high status and better performance on the self-promotion gender gap is thus ambiguous. We ask: \emph{How does the gender gap vary with the academic status of the scholar and the performance of their work?}

Lastly, self-promotion is a strategic action that likely depends on expected returns, such as gains in visibility. If expected returns vary by gender, scholars are likely to engage in it differentially. Accordingly, \emph{Is there a gender gap in the returns for self-promotion?} 
Our analyses are associational and only suggestive of causality, a point we return to in the discussion. 

By quantifying the self-promotion gap over multiple years and at the scale of English-language scientific literature covering all fields, we provide a much-needed comprehensive empirical examination of a key mechanism behind the lower visibility of women's scientific achievements. 
Our findings point to a gender gap that appears in all fields and is the most pronounced for productive women from top-ranked institutions who publish in high-impact journals. 
Complemented by our finding that women's self-promotion is associated with lower returns (in terms of social media mentions), our work highlights the critical need to rethink and improve existing practices in promoting and evaluating scholarly work.

\section{Materials and Methods}

We explore these questions using a multi-disciplinary dataset of 2,834,829 research papers published between 2013 and 2018 by 3,503,674 unique authors. These papers have received at least one mention on some online platform as tracked by Altmetric.com \cite{alt}. Altmetric is the most comprehensive service to date for monitoring online posts about research papers.
Our self-promotion and online visibility measures focus on Twitter (now ``X''). During the time frame studied here, Twitter was the most commonly used platform for online science dissemination, accounting for 92\% of all paper mentions on social media platforms \cite{peng2022dynamics}. 
Our dataset combines the Altmetric data on papers' Twitter mentions with Microsoft Academic Graph \cite{wang2019review}, which provides rich metadata on each author and paper, including affiliations, citations, and research fields (see details in \sm{Data collection}).

\begin{figure*}[ht!] 
\centering
\includegraphics[trim=0mm 0mm 0mm 0mm, width=0.7\linewidth]{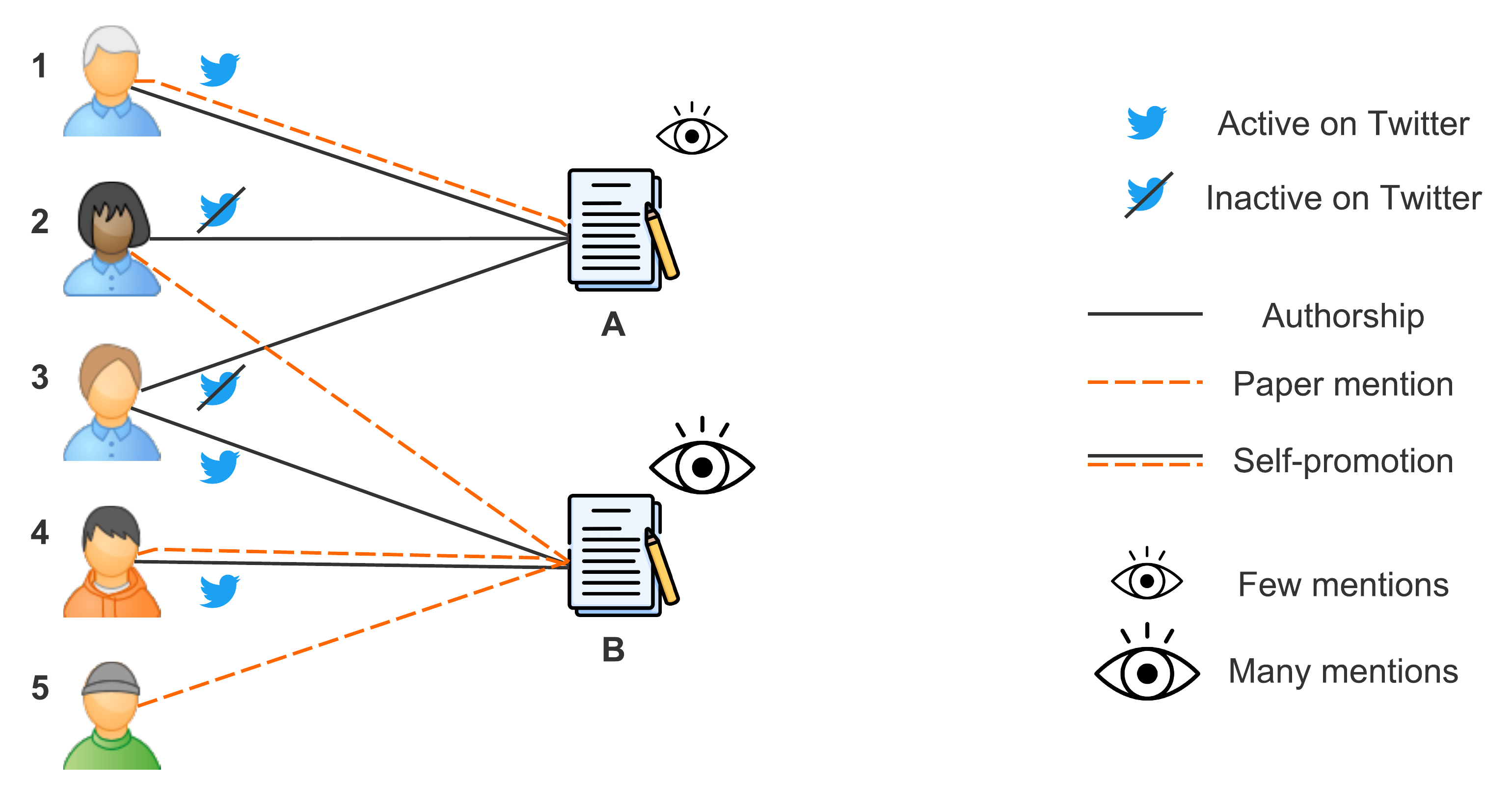}
\caption{\textbf{Illustration of the studied process.} A paper can have multiple authors, and it can receive several mentions from different Twitter users including academic authors and non-academics. We treat each (paper, author) pair, or each authorship shown in this plot, as the unit of analysis, and accordingly code whether the author is active on Twitter at the publication of the paper (the Twitter icon). In this illustration, Person 1, who is active on Twitter at the time of Paper A's publication, self-promotes. Neither of his co-authors, Person 2 or 3, are active on Twitter and therefore do not self-promote. By the time Paper B is published, Person 2 is active on Twitter and mentions the paper, although she is not one of its authors. Out of Paper B's authors, Person 4 self-promotes, but Person 3 does not, despite that he now becomes active on Twitter. Paper B is also mentioned by Person 5, who is not an academic. Among scholarly authors, we distinguish women and men to investigate the likelihood of self-promotion for each (paper, author) pair. Furthermore, for each paper, we record the number of mentions on Twitter, which we use to examine potential gains in visibility associated with self-promotion.
}
\label{fig-one}
\end{figure*}

Because authors might promote some of their papers but not others, and not all authors of a paper are equally likely to self-promote it, our unit of analysis is the (paper, author) pair. We identify 11,396,752 (paper, author) pairs, each of which is a candidate for self-promotion (\fref{fig-one}).
This design enables us to isolate the role of gender neatly because we can fully account for differences in papers and authors that can affect self-promotion. For example, the same author may consider some topics or publications as more worthy of self-promotion, and different authors of the same paper may base individual self-promotion decisions on their authorship roles. 

To detect whether an author self-promotes a paper, we design and validate a heuristic-based method to match author names to the list of Twitter usernames that have mentioned the paper, achieving an out-of-sample $F1 > 0.95$ (\sm{Detecting self-promotion}).

Additionally, for each (paper, author) pair, we estimate whether the author was active in tweeting academic papers when the paper was published (\fref{fig-one}). 
While no method can perfectly match all scholars to their Twitter accounts because they can adopt usernames unrelated to their academic names and provide little or no information on their Twitter profile, we use a validated, state-of-the-art external dataset that achieves high accuracy in this setting \cite{mongeon2023open, costas2020large} (\sm{Establishing active status on academic Twitter}).

We infer author's gender using first names \cite{ford2017someone}. Out of 3,503,674 unique authors, 58\% are inferred to be men and 42\% women (results with gender-neutral names included as a separate category are shown in \sm{Table S4}). We validate our gender inferences in two ways. First, we compare the algorithm's inferences to an auxiliary dataset containing the self-reported gender of 432,888 physical scientists. 
Second, we compare inferences to the manually labeled gender of 100 randomly selected authors in our data. These validations reveal that the gender classifier is very accurate with the $F1$ score being close to 0.9 (\sm{Gender inference}).
At the level of (paper, author) pairs, men account for a substantially larger fraction than women (64.7\% vs 35.3\%).

Our first set of models investigates gender differences in the likelihood of self-promotion. The outcome variable ``self-promoted'' is an indicator of whether the author in an (paper, author) pair mentioned the paper on Twitter. To fully account for differences between papers, we use a mixed-effects logit model with random effects for each paper. This setup enables us to compare self-promotion by different authors of the same paper. The model includes gender and a set of covariates that are likely to affect self-promotion. The paper-level controls include the publication year, the impact factor of the journal publishing the paper, and the research fields associated with the paper. In terms of author-related factors, we include the number of authors, authorship position, affiliation location and rank, and author's previous publication and citation counts. 
Controlling for these covaritates, which themselves might be related to gender, means that our models estimate gender's relatively \textit{direct} association with self-promotion instead of its \textit{overall }association (\sm{Modelling the likelihood of self-promotion}).
Additionally, we use propensity score matching to test the robustness of our results. This technique allows us to identify (paper, author) pairs that are similar on all observed covariates except the gender of the author (see details in \sm{Robustness check using propensity score matching}).

Our second set of models examines the return on self-promotion in terms of visibility. 
We use a negative binomial regression \cite{hilbe2011negative} to model a paper's total number of Tweet mentions as a proxy for its online visibility. The model includes an interaction term between author gender and self-promotion to examine how self-promotion is associated with tweet mentions differently for men and women.
The unit of analysis is still each (paper, author) pair so that we can measure visibility premium across different author characteristics, even for authors of the same paper. 
Crucially, we focus this analysis on the subset of observations where the author was active on Twitter when the paper was published. This enables us to include the control for author's log-scaled number of followers on Twitter in addition to the controls defined previously.

\section{Results}

\paragraph{The Increasing Gender Gap in Self-Promotion.}

\begin{figure*}[ht!] 
\centering
\includegraphics[trim=0mm 0mm 0mm 0mm, width=0.95\linewidth]{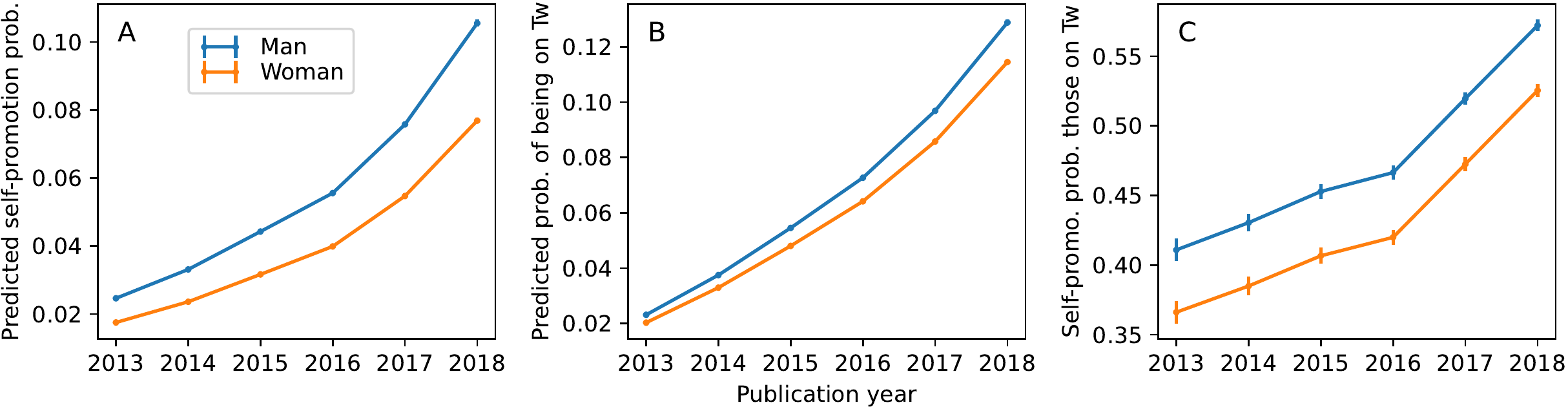}
\caption{\textbf{The gender gap in self-promotion is not fully explained by women's under-representation on Twitter}. \textbf{A}, Predicted probability of self-promotion by gender based on a mixed-effects logistic regression model fitted to 11,396,752 (paper, author) pairs where the dependent variable is coded as 1 if the author has self-promoted the paper. \textbf{B}, Predicted probability of active presence on Twitter by gender, based on a regression model with the same controls but where the dependent variable is coded as 1 if the author is active on Twitter at the paper's publication date. \textbf{C}, Predicted probability of self-promotion for the subset of 618,742 (paper, author) pairs where the author is active on Twitter.
All three regression models have the same controls, including publication year, journal impact factor, authorship position, number of authors, affiliation rank and location, author productivity and number of citations, research topics, and paper random effects. Control variables have been set to their median values to create these plots.
Error bars indicate 95\% bootstrapped confidence intervals.}
\label{fig-two}
\end{figure*} 

\fref{fig-two}A shows the predicted self-promotion probability by year for men and women (see regression details in \sm{Table S1}). 
Male authors had a 2.5\% - 10.6\% chance of self-promoting papers between 2013 and 2018, while comparable female authors had a 1.8\% - 7.7\% chance of self-promoting, which is about 28\% lower on average. 
This disparity exists even among female and male authors of the same paper.
Critically, while both genders self-promote more often over time, we find that the gender gap has been growing substantially. 
The difference does not disappear when measuring self-promotion as only original tweets or retweets (\sm{Fig. S3}), suggesting that the gap exists even in the arguably less direct self-promotion act of retweeting.

Since the gender gap in self-promotion could be impacted by potential biases in gender inference or group dynamics within mixed-gender teams, we perform a suite of robustness checks. The gender gap is remarkably consistent when (1) including author names classified as gender-neutral as a separate category (\sm{Table S4}), (2) excluding authors names associated with East Asian ethnicities that usually do not encode a clear gender signal (\sm{Table S5}), (3) considering self-promotion as being the first author to tweet about the paper (\sm{Table S6}), and (4) subsetting the data to solo-authored papers only (\sm{Table S7}).
These additional analyses show that authors with gender-neutral names self-promote less often than those with gender-distinct names, the gender gap remains after excluding East Asian ethnicities, and potential role assignments do not drive the self-promotion gap in mixed-gender teams.
As a final test, we use propensity score matching to reduce the confounding effects that may be unaccounted for by the regression model. 
We still find that women self-promote their research less often than comparable men (\sm{Table S8-S9}).

\paragraph{The Gender Gap is Not Explained Entirely by Gender Differences in Twitter Adoption.}

In our previous analysis, we do not distinguish between authors who do not self-promote because they are not active on Twitter and those who are active but choose not to post about their papers. However, women's lower self-promotion on Twitter may be caused by their choices conditional on being active on Twitter, or simply adopting Twitter at lower rates \cite{zhu2019gender, costas2020large}.
To explore the latter mechanism, we first run a regression model with the same specification as for predicting self-promotion but changing the dependent variable to whether the author is active on Twitter at the time of the paper's publication (see regression details in \sm{Table S2}).
\fref{fig-two}B shows that among all authors, men are indeed more likely than comparable women to be active on Twitter. 
Running the regression from \fref{fig-two}A on the subset of (paper, author) pairs where the author was active on Twitter at the time of publication (see regression details in \sm{Table S3}) shows a considerably higher level of self-promotion among Twitter-active authors than among all authors (\fref{fig-two}C). Yet, even in this subset, women self-promote 9.4\% less than men, suggesting that the substantial gender gap is not fully explained by differences in active use of Twitter. 

\paragraph{The Gap is Robust to Variation in an Area's Gender Representation.}

\begin{figure*}[ht!] 
\centering
\includegraphics[trim=0mm 0mm 0mm 0mm, width=\linewidth]{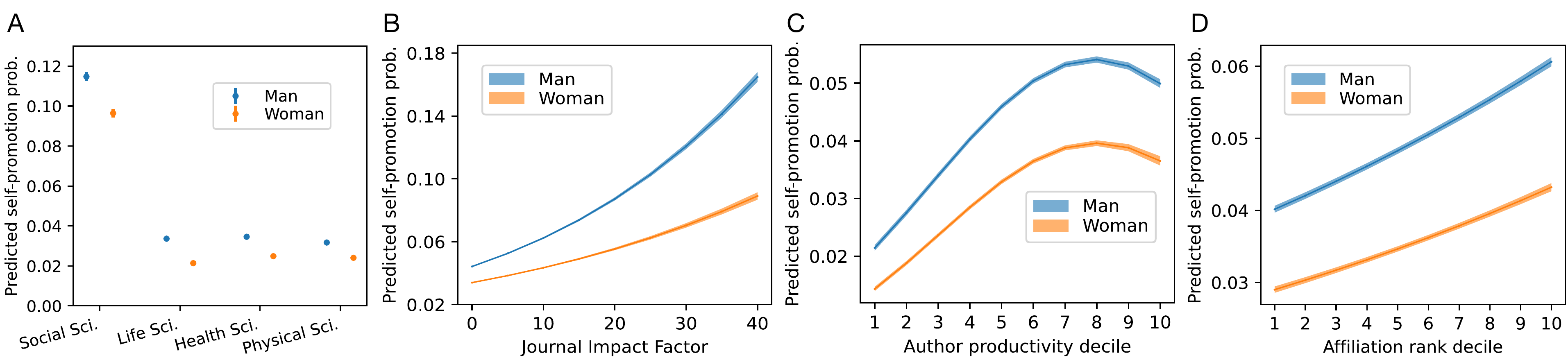}
\caption{\textbf{The gender gap in self-promotion is similar across four broad areas but increases with higher performance and status quantified by journal impact factor, author productivity, and affiliation rank.}
\textbf{A}, The predicted gender gap in self-promotion is of similar magnitude across all four broad research areas. We fit a separate model for each area while still controlling for fine-grained subject areas. 
\textbf{B-D}, The likelihood of self-promotion as a function of journal impact factor, author productivity, and affiliation rank, predicted based on our most comprehensive model (Model 5 in \sm{Table S1}) that also adds an interaction term between gender and the corresponding variable.
To quantify productivity, we decile author's number of publications as a measure of their productivity (\textbf{C}, a larger bin indicates a more productive decile). Similarly, we decile author's affiliation rank (\textbf{D}, a larger bin indicates a higher rank decile).
All predictions are based on setting control variables at their median values in our data.
Error bars indicate 95\% bootstrapped confidence intervals.}
\label{fig-three}
\end{figure*} 

Does a higher representation of women in a given research area decrease the gender gap in self-promotion?
To answer this question, we estimate the overall gender gap in self-promotion across the four broad research areas of social, life, health, and physical sciences. In our dataset, the authorship share of women for papers from these individual areas is 46\%, 43\%, 44\%, and 28\%, respectively.
\fref{fig-three}A shows the predicted self-promotion probability, which is higher for men than for women in all four broad areas (see regression coefficients in \sm{Table S10-S11}).
Social scientists are, on average, about three times as likely to promote their research as scientists in the other three areas (0.10 vs. 0.03). Yet, the gender gap is comparable across research areas regardless of gender balance. Even in physical sciences, where only approximately 1 in 4 authors are women, the gap is comparable to the gap in broad areas that are approaching parity in authorship in our data.
This suggests that the gender gap in self-promotion is consistently large, even in the presence of substantial variation in an area's gender representation.

\paragraph{The Gender Gap is Largest for High-performing, High-status Women.}

To examine the association between the self-promotion gap and a widely used scholarly performance measure, we estimate self-promotion rates across different levels of journal impact factor for both genders (\fref{fig-three}B).
As expected, the overall level of self-promotion is positively associated with journal impact factor. 
However, the gender gap is larger for higher impact factor journals. In particular, for papers published in journals with an impact factor of 40, men are 78\% more likely to self-promote than women (0.16 vs 0.09), whereas, for low-impact-factor publications (impact factor below 5), the predicted self-promotion probability for men is only 33\% higher than for women (0.04 vs 0.03). 

To investigate the effect of status, we focus on the author's academic standing as quantified by productivity and affiliation rank. We show in \fref{fig-three}C the predicted self-promotion probability as a function of the author's number of publications, which is an established measure of their productivity.
Regardless of gender, a scholar's self-promotion probability increases as their research experience grows and then saturates in the mid-to-late career stage.
One possible reason is that junior researchers spend most of the time conducting research to build their portfolio, while the task of marketing papers is often taken on by more experienced scholars \cite{wren2007write, sekara2018chaperone}.
However, this transitional change does not apply to men and women equally.
\fref{fig-three}C shows a larger gender disparity among more experienced scholars: the predicted difference in self-promotion between men and women is 0.014 vs. 0.007 for authors in the most vs. the least prolific decile.

We repeat the analysis across affiliations of different ranks as another proxy for the author's status (\fref{fig-three}D). 
We find that authors of both genders from the highest-rank decile are about 43\% more likely to self-promote than those from the lowest-rank affiliations. 
Moreover, as with productivity, the gender gap is larger for more prominent affiliations: the predicted self-promotion difference between men and women is 0.02 vs. 0.01 for authors in the highest vs. the lowest affiliation rank decile. 

These results suggest that there are universal and well-defined self-promotion patterns, which are associated with performance (journal impact factor) and status (research productivity and affiliation rank). 
However, men and women's self-promotion changes with different intensities in response to these mediating factors, resulting in an even larger gender gap at the high-performing and high-status end of the spectrum.
This evidence points to the potential role of pushback against high-achieving women and calls for more investigations to establish a causal relationship.

\paragraph{Differential Return on Self-Promotion.}

\begin{figure*}[ht!] 
\centering
\includegraphics[trim=0mm 0mm 0mm 0mm, width=0.4\linewidth]{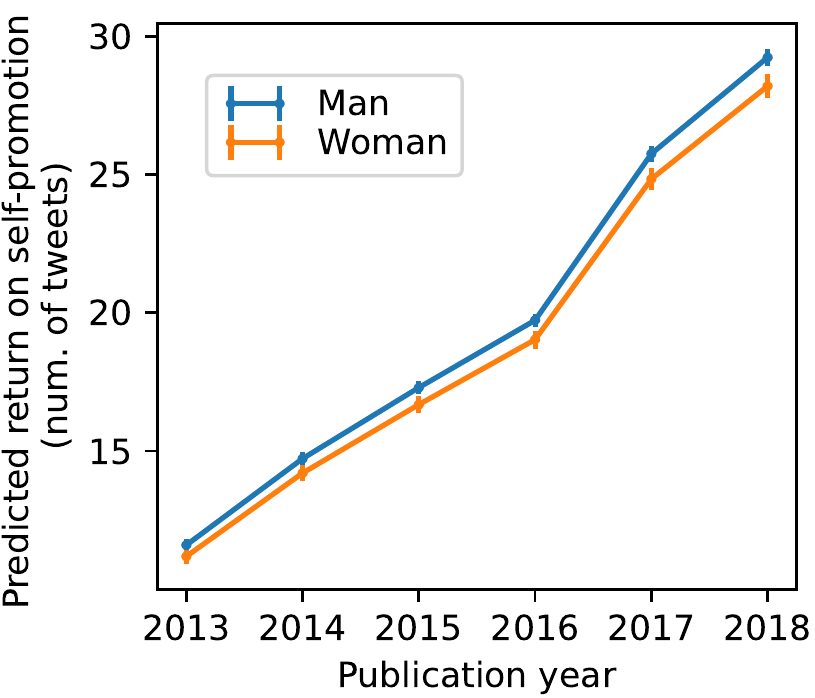}
\caption{\textbf{Women's self-promotion is associated with fewer tweet mentions than men's.} 
We show the yearly marginal effect of self-promotion for both genders, based on a negative binomial regression fitted to 618,742 (paper, author) pairs where the author was active on Twitter at the paper's publication date. 
The model additionally controls for author's log-transformed follower count and adds an interaction term between author gender and self-promotion status. 
Predictions are based on setting all control variables at their median values in the fitting data.
Error bars indicate 95\% bootstrapped confidence intervals.}
\label{fig-four}
\end{figure*} 

A critical closing piece in our longitudinal investigation of scholarly self-promotion on social media examines the visibility gains from promoting one's own paper (\fref{fig-four}).
As expected, self-promotion is strongly associated with our measure of visibility (i.e., a paper's total number of tweet mentions), and papers have been receiving more and more mentions over the studied time period. Importantly, however, women's self-promotion is associated with fewer mentions than men's (see regression coefficients in \sm{Table S12}). 

To test the robustness of this finding, we run four additional tests. 
First, we confirm that the result is consistent when fitting a separate model for papers published in individual years (\sm{Table S13}).
Second, to isolate the gender effect, we eliminate potential self-promotion dynamics in mixed-gender teams by fitting a separate model for solo-authored papers. We find that the results are qualitatively similar (\sm{Table S14}).
Third, we show that the finding is robust when defining self-promotion as sharing the paper on social media within one day of publication (\sm{Table S15}). 
This additional analysis shows that the gender difference in return equally applies to early self-promotion, indicating that it is unlikely that women's self-promotion is ineffective because they don't do it in a timely manner.
Fourth, the result also holds when we consider mentions of the paper by only scientists or non-scientists (\sm{Tables S16-S17}), indicating that neither of these groups alone can account for the gender difference in return on self-promotion.

\section{Discussion}

Based on the analysis of 3,503,674 authors and their 2,834,829 research papers published between 2013 and 2018 across all fields of scholarship, we find a large gender gap in self-promotion on Twitter.
Women are, on average, 28\% less likely than men to self-promote their papers even after taking into account important confounding factors such as the year of publication, fine-grained research fields, as well as author- and paper-level characteristics related to status and performance. 
The gap even appears when comparing authors of the same paper and increases over the span of six years studied here.
Crucially, differential adoption and usage of Twitter do not explain the gap. While our results show that women are slightly underrepresented on academic Twitter relative to men, the gender gap in self-promotion appears even among those active on Twitter. 
The gender gap in self-promotion is consistent with gender gaps observed offline, such as in self-citation \cite{king2017men, andersen2019meta} and embellished presentation of results \cite{lerchenmueller2019gender}.

Delving into factors that can moderate the size of the gender gap, we find that it is \emph{not} substantially smaller in more gender-balanced fields. This suggests that improved gender representation is unlikely to fix the self-promotion gap in the short term. 
Importantly, we also find that factors signaling status and performance can substantially moderate the size of the gender gap. 
In particular, the gap in self-promotion grows for papers published in higher-impact journals and when authors are more productive and come from higher-ranked institutions.
These results suggest that women's concerns about pushback in response to their self-promotion may dominate the likely high self-assessment associated with being an established scholar and assurance that the contribution is worth promoting \cite{ma2019women, shor2022women}. When high-achieving women self-promote less than high-achieving men in a scientific workforce characterized by a leaky pipeline, where there are fewer and fewer women in higher ranks, women's visibility is expected to remain low.

We also find a gender gap in the visibility return associated with self-promotion.
Although self-promotion is strongly linked to higher engagement for both genders (reiterating its importance), the increase in mentions is statistically smaller for women than for men.
Overall, our observational study provides the crucial first step in documenting surprising and consequential nuances of a substantial gender gap in self-promotion across all areas of science.
 
The study is not without limitations.
First, we study a single social media platform. 
Although Twitter (now ``X'') is the largest platform based on mentions of scientific papers, it would be interesting to extend the investigation to other social media sites. 
Second, we use name-inferred gender as a proxy of the author's gender identity. 
Our two validations show that, overall, there is a high degree of agreement between the predicted and manually verified or self-reported genders. However, the errors are not randomly distributed. In particular, we observe that the accuracy is lower for authors from East Asian countries \cite{lockhart2023name}. 
Nevertheless, our main result is similar for authors with non-East Asian names.
We also find that including authors with gender-neutral names in the analysis shows that these scholars are the least likely to self-promote. Our observation calls for further research to examine the composition of this group and the specific challenges they face in self-promotional contexts.
Another limitation of our analysis is that authors of non-binary gender are not identified, which we hope will be rectified by future work. 

Third, our study requires matching authors to Twitter users to detect self-promotion. We address this problem using a name-based matching algorithm. Although our validation shows that our matching process achieves high accuracy, the algorithm is not without errors. Future work can develop more sophisticated approaches to address this issue.
Finally, we rely on an open dataset of scholars identified on Twitter to detect if they were active on the platform at the publication date of a paper. The external dataset does well on our validation tests, yet it cannot identify all authors who were active on Twitter during the entirety of the six years we investigate.

Despite these limitations, our study offers novel insights into scholars' self-promotion on social media, enriching our understanding of the broader gender inequity in scientists' online visibility.
As one policy implication, our work informs the way we interpret online visibility metrics. Lower visibility metrics for women are a systemic issue with complex underpinnings and are explained at least in part by men's higher enthusiasm and greater returns for self-promotion. This should be considered when using such metrics in research evaluation. 
Another ramification of these findings is that self-promotion is associated with greater engagement. If this link is causal, institutions should provide training, dedicated time, interventions, and support in case of online harassment to facilitate self-promotion for everyone engaging in it.

Our study also opens up clear directions for future research. 
We hope it will encourage examinations of the mechanisms causing the gender gap in self-promotion and its returns. Such investigations would be critical because algorithms on online platforms may amplify small initial differences, as produced via differential self-promotion, into large disparities in ultimate attention \cite{lin2023rich}. 
Such gender differences in online visibility may then result in inequalities in conventional markers of scientific excellence like publications and citations \cite{milojevic2020towards, bardus2020use, luc2021trends}. 
Additionally, future work might directly examine why the gender gap is substantial even in relatively gender-balanced fields, and why it is larger when promoting achievements widely considered valuable, such as papers in high-impact-factor journals.

\begin{addendum} 
    \item[Data Availability] The Altmetric data can be accessed free of charge by researchers from \url{https://www.altmetric.com/research-access/}. The Microsoft Academic Graph is available at \url{https://www.microsoft.com/en-us/research/project/open-academic-graph/}. The journal impact data are from Web of Science. A public repository of our data is available at \url{https://doi.org/10.6084/m9.figshare.21843330}. Our code can be accessed at \url{https://github.com/haoopeng/gender_promotion}.
    \item[Acknowledgements] We thank Altmetric, IOP Publishing, and Microsoft Academic Graph for sharing the data. The authors thank Orsolya V\'as\'arhelyi and Bogdan Vasilescu for helpful discussion. This work has been partially funded by NSF CAREER Grant No IIS-1943506 and by the Air Force Office of Scientific Research under award number FA9550-19-1-0029.
\end{addendum}

\clearpage

\bibliography{references}

\end{document}


\title{The Gender Gap in Scholarly Self-Promotion on Social Media \\ (Supplemental Materials)}

\author{Hao Peng, Misha Teplitskiy, Daniel M. Romero, Em\H oke-\'Agnes Horv\'at}

\date{\today}

\maketitle

\section{Supplementary Text}

\subsection{Data collection}


\textbf{Altmetric.} 
Our data are based on the most complete record of research papers' online mentions, maintained by Altmetric.com \cite{alt}.
This service has been tracking the online mentions of research outputs since 2011 on different platforms, including news media and social media such as Twitter and Facebook. 
We accessed the database (referred to as ``Altmetric'' hereafter) on October 8, 2019.
Altmetric matches online attention to papers based on their unique identifiers, such as the Digital Object Identifier (DOI), PubMed ID, and arXiv ID.
Utilizing these identifiers, it also collapses the attention of different versions of each paper into a single unique record \cite{alt2021how}. This ensures that the data contain the complete mentions of each paper.
We obtain all the mentions of research papers published between 2013 and 2018 in the largest platform, Twitter, which consists of about 74\% of all mentions (posts) and 92\% of all social media mentions in the entire database \cite{peng2022dynamics}.

Due to Altmetric's data license agreement with Twitter, the dataset contains only the tweet ID for each tweet. We thus collect all tweets using the Twitter API. 
Due to account deletion or changes in the privacy settings by some Twitter users, we have successfully retrieved about 90\% of all tweets. 
The Altmetric data also provide metadata for each paper, such as the DOI, publication year, publication venue, and research topics.
The disciplinary catalog uses 26 Scopus Subject Areas, which belong to four broad disciplines: Social Sciences, Life Sciences, Physical Sciences, and Health Sciences \cite{scopus}. The classification is performed by in-house experts based on the aim and scope of the content a journal publishes \cite{scopus}.\\

\textbf{Microsoft Academic Graph.} 
We use the Microsoft Academic Graph (MAG) database \cite{sinha2015overview, wang2019review} (accessed on June 01, 2019) to retrieve other metadata for each paper based on the DOI. 
We obtain essential author information for each paper, including author's name and affiliation, affiliation location and rank.
We also consider the authors' previous number of citations and publications up to the publication year of each paper, which reflects authors' previous productivity when deciding whether to self-promote or not, assuming authors tend to self-promote soon after the publication of a paper.

MAG leverages data mining and artificial intelligence techniques to address author conflation and disambiguation, which ensures that the author's number of publications is counted accurately \cite{wang2019review, mag2018aut}.
Our dataset at this stage consists of 14,521,585 (paper, author) pairs for 2,983,706 papers after ensuring that we have complete metadata.

\subsection{Gender inference}

\textbf{Gender prediction.}
We use Ford's algorithm \cite{ford2017someone} to infer the gender of an author based on their first name.
The algorithm returns, for a given name, one of 4 categories: Woman, Man, Unisex, Unknown. 
A name is predicted as ``Woman'' (``Man'') if it is used for women (men) at least twice as frequently as for men (women), based on data from national statistics institutes \cite{gendercomputer}. Otherwise, the name is labeled as ``Unisex'' (gender-neutral). If the name is not found in the database, it is labeled as ``Unknown.''

The percentage of 14.5M observations for Man, Woman, Unknown, and Unisex is 51\%, 28\%, 16\%, 5\%, respectively. 
We exclude ``Unisex'' and ``Unknown'' categories from our analytical sample, which consists of 11,396,752 (paper, author) pairs for 2,834,829 papers by 3,503,674 unique authors (58\% Man vs. 42\% Woman).
We treat ``Man'' as the reference category in all regressions.\\

\textbf{Agreement with manually assigned gender labels.}
We evaluate the performance of this algorithm based on a random sample of 100 authors in our data, for which we manually label their gender.
In the labeling process, the author's gender is determined based on their pronouns and profile pictures displayed on their personal websites, institutional directories, and Wikipedia pages found via Google searches of author names. 
We use the author's affiliation to disambiguate multiple authors with the same name. 
If the gender could not be verified, it is labeled as ``Unknown.''
There are 19 women, 57 men, and 24 authors with unknown gender in the labeled sample.
Based on the set of 76 authors with confirmed gender, the algorithm achieves an accuracy of F1$_m$ = 0.91 and F1$_f$ = 0.88.
A different algorithm, the Genderize API \cite{genderize, lerchenmueller2019gender, coding2015, karimi2016inferring}, produces a similar result.\\

\textbf{Validation with self-reported gender.}
We also validate the accuracy of the gender prediction using author self-identified gender labels in the manuscript submission data provided by IOP Publishing (\url{https://ioppublishing.org/}), an academic publishing company specializing in the field of physics.

Each author has self-reported their gender and country of residence, such as China, India, U.S., Canada, Australia, etc.
In our evaluation, we focus on authors with self-reported gender as either man or woman and whose names are predicted as either man or woman, as our actual analysis is focused on men and women.
There are 432,888 authors, submitting to 62 journals, in this data. 
Here, we list authors from China (the largest single group) as a separate group because Chinese names typically do not encode a clear gender signal when written in latin characters and different Chinese names can have the same latin characters \cite{jia2019gender}. For instance, we find that the vast majority of names (475 such names in total) with both man and woman as self-reported gender labels are from China. 
The prediction accuracy for Chinese names is lower than non-Chinese names, as expected, but the overall F1 score is close to 0.9 (Table S18).\\

\textbf{Robustness check on non-East Asian names.}
Since the gender prediction is typically less accurate for East Asian names \cite{santamaria2018comparison}, we repeat our analysis by excluding observations whose author names are predicted to be East Asian ethnicities using the Ethnea API  \cite{torvik2016ethnea}.
For a specific name, \textit{Ethnea} assigns the ethnicity probabilities based on matched authors in the PubMed database.
In the case of two or more predicted ethnicities, we take the one with the highest probability. 
\textit{Ethnea} predicts 26 individual ethnicities. We categorize seven of them as East Asian, including \textit{Chinese}, \textit{Indonesian}, \textit{Japanese}, \textit{Korean}, \textit{Mongolian}, \textit{Thai}, \textit{Vietnamese}.
We obtain consistent results on the subset of the data with non-East Asian names (Table S5). 

\subsection{Detecting self-promotion}

Each paper has a list of tweets that mention the paper, with each tweet containing the Twitter handle and the screen name of the user (referred to as ``tweet names'' hereafter). 
We define self-promotion as the author posting a tweet sharing the unique identifier of their paper, such as the DOI, PubMedID, arXiv ID, etc.
Self-promotion on Twitter comes in two different forms, which can manifest in the type of tweet that shares the research, including (i) an original tweet, or (ii) a retweet.
We construct a binary dependent variable indicating whether the author self-promoted the paper or not, based on both retweets and original tweets.
The two types of self-promotion differ in how direct the promotion is: the original-tweet-based promotion comes from the authors themselves, whereas the retweet-based promotion originates from others (e.g., an author A retweets a tweet from others sharing A's paper). 
Self-promotion in the form of posting an original tweet potentially indicates a stronger intention to advertise one's paper than does posting a retweet, which is a more indirect promotion. 

We determine if an author is among the users who have tweeted the paper based on string matching.
There is no perfect method to match author names with tweet names because scholars can use any string as their handle or screen name on Twitter. 
We thus adopt a simple ``containment-matching'' approach that searches the author name in tweet names---if either the first name or the last name string is contained in the user handle or screen name (lowercased), we consider the user to be the author of the paper.
In case of multiple matches, we use the one with the highest fuzzy matching score \cite{fuzzy}.

We validate this method using a random sample of 100 papers (each with at least one tweet mentioning it) with manual verification. Due to having multiple authors per paper, there are 521 (paper, author) pairs in the manual labeling process. For each observation, we verify the author against all tweets of the paper to check if the author is among the tweet users. 
This method achieves an initial F1 score of 0.85 (precision: 0.77 and recall: 0.95).



We refine this method through an iterative process by experimenting with various new heuristics to address false predictions and evaluating its performance based on an independent manual verification process at each iteration.
In the final version, we use ``containment-matching'' only if the tweet names are single-token strings and the author's first name (or last name) has at least four characters; otherwise, we use ``token-matching'': i.e., the first name or the last name should be matched to the tokens of tweet names (split by space or underscore). 
This heuristic achieves an out-of-sample precision of 0.95 and a recall of 1.00 (F1 score: 0.97).


\subsection{Establishing active status on academic Twitter}

The crucial practical challenge in establishing which scholars are active on Twitter is to match them to their Twitter accounts reliably. This matching is difficult because researchers often use Twitter handles unrelated to the names they use for academic papers. 
Moreover, even if all authors use Twitter handles containing parts of their names, it would still be computationally challenging to match author names to all possible Twitter handles currently in use. 

We conduct a thorough literature search to identify the latest methods for linking academics to Twitter users. While no method can perfectly identify whether an author has a Twitter account, the state-of-the-art approach by an independent research team using open datasets produces valuable results \cite{mongeon2023open, costas2020large}. 
This method utilizes a heuristic to match authors in OpenAlex \cite{priem2022openalex} to a massive dataset of Twitter users who have mentioned a research paper in the Crossref Event data \cite{crossref}. 
Thus authors identified in this external dataset are at least somewhat ``active'' on Twitter. 
Although the active authors identified in this dataset might still be different from the set of all active authors (e.g., authors whose Twitter usernames are entirely unrelated to their author names), it is a high-precision indicator of which scholars are active on Twitter.

Our data span six years, and authors could have joined Twitter at any point during this timeframe and published multiple papers. With this in mind, we have to carefully code the active status on Twitter for each (paper, author) observation.
Specifically, a (paper, author) pair is coded as active on Twitter if two conditions are met: (1) the author is identified as having an active Twitter account in the external dataset \cite{mongeon2023open}, and (2) the paper's publication date is after the author's earliest tweet time in our Altmetric dataset.

\subsection{Modeling the likelihood of self-promotion}

We use a mixed-effects logistic regression \cite{bates2015fitting} to estimate the probability that an author self-promotes a paper as a function of their inferred gender while controlling for the following variables.

\textit{Year of publication}: Scholars are increasingly promoting their research on Twitter over time (Fig. S1), thus it is important to control for the publication year of a paper.

\textit{Journal impact factor}: Papers published in high impact journals may be more likely to be shared by their authors (Fig. S2). We obtain the journal impact factor from The Web of Science (2018 version).

\textit{Author's affiliation rank}: Authors from higher-ranked institutions may be more likely to self-promote (Fig. S2). We thus consider the rank of their affiliations provided in the MAG. When an author has multiple affiliations in a paper, we use the one with the highest rank. 
This ranking metric estimates the relative importance of institutions using paper-level features derived from a heterogeneous citation network; while similar to h-index, the method has been shown to produce more fine-grained and robust measurement of impact and prestige \cite{sinha2015overview, wang2019review}. 
To smooth this variable, we categorize the rank values into ten equally-sized bins (note that a smaller bin indicates a higher rank category).

\textit{Author's affiliation location}: We infer the country of author's affiliation using the latitude and longitude provided in the MAG. 
We classify countries into two categories: (1) U.S., (2) international (non-U.S.). 
For authors with multiple affiliations, we only classify them as ``U.S.'' when at least one of their affiliations is located in the U.S.
We use ``U.S.'' as the baseline to control for the fact that Twitter is an American social media platform that is more likely to be adopted by authors based in the U.S.

\textit{Authors' previous number of publications}: Authors' academic standing or research experience can influence their likelihood to self-promote. To measure this variable, we count each author's total number of publications before the publication year of the target paper based on MAG. We also categorize this numerical variable into ten equally-sized bins to reduce noise and outliers.
Since there is a quadratic relationship between author productivity and self-promotion probability in the data (Fig. S2), we include a second order polynomial term for author productivity in the model.

\textit{Authors' previous number of citations}: Authors' citation counts can impact both their likelihood of self-promotion and the return associated with self-promotion. We count each author's total number of citations before the publication of the target paper using the MAG database. We log-transform this highly skewed variable to make it more interpretable.

\textit{Number of authors}: We count the number of authors in each paper. Having more coauthors in a paper may negatively impact an author's likelihood to promote the paper on their own.

\textit{Authorship position}: Different authors often play different roles in multi-author projects.
For example, authors who play a supportive role in the project may self-promote less frequently (Fig. S1). This variation is often captured by the authorship position in the paper. We thus control for the position of an author using a dummy variable with four categories: (1) first position, (2) middle position, (3) last position, (4) solo author. The ``last position'' is used as the reference category in the regressions. Note that when ``solo author'' = 1, the other three positions are all coded as 0. 

\textit{Research fields}: Not all scholars employ Twitter as a channel to share their research, and scientists' representation on Twitter varies across disciplines \cite{ke2017systematic}. To control for field-specific effects, we use the 26 Scopus Subject Areas. Each subject area is treated as a fixed variable in the regression, and its value is coded as 1 if the paper is assigned that subject (0 otherwise). Note that a paper can belong to multiple subject areas.

\textit{Paper random effects}: Individual papers have different degrees of newsworthiness (e.g., biomedical papers have much more online coverage than papers from other disciplines \cite{banshal2019disciplinary}). Different papers may vary in the likelihood of being shared on social media by their authors. Gender representation online also varies across disciplines \cite{vasarhelyi2021gender}. To capture such paper-level variations, we add random effects for each paper in the model.

\subsection{Robustness check using propensity score matching}

As a robustness check, we use propensity score matching to reduce the confounding effects that may be unaccounted for by the logistic regression model.
We estimate the propensity score for each observation using a logistic regression with the author's gender as the dependent variable and all other confounding factors defined previously as controls.

We then match each woman observation to one man observation using the nearest neighbor matching based on the propensity score. In this way, we construct a sample of man observations that is comparable on all observed covariates to the population of woman observations (Table S8).
Using the matched sample, we still find that women self-promote their research less often than comparable men (Table S9).

\subsection{Predicting the paper's number of tweet mentions}

We use a negative binomial regression to predict the total number of tweets of a paper. The unit of analysis is still each (paper, author) pair. We focus this analysis on the subset of observations where the author is active on Twitter at the publication date of the paper. The key independent variables are gender, self-promotion status (binary), and their interaction effect. 
The control variables are the same as defined previously, and include the author's number of followers on Twitter (log-scale). 
Due to computational cost and time complexity, the paper random effects are not included in this model.

\section{Supplementary Figures}





\begin{figure*}[ht!] 
\centering
\includegraphics[trim=0mm 0mm 0mm 0mm, width=\linewidth]{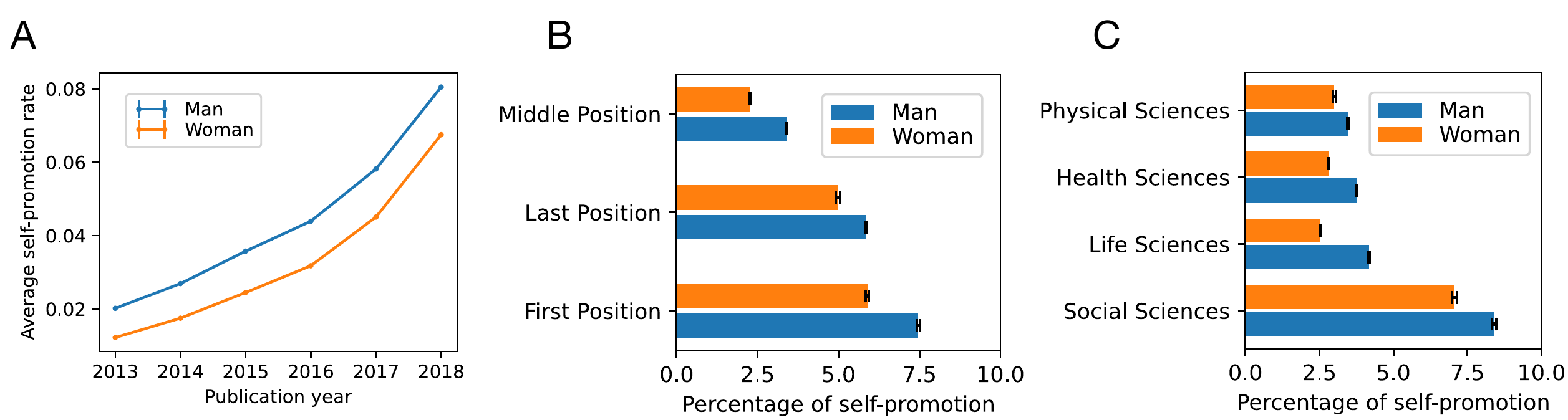}
\caption{The average self-promotion probability by gender as a function of publication year, authorship position, and broad discipline in our data. \textbf{A}, The raw self-promotion probability quadruples from 2013 to 2018 for both genders. \textbf{B}, The percentage of self-promotion grouped by authorship position. \textbf{C}, the same as \textbf{B}, but for the breakdown by papers' discipline. Confidence intervals represent 95\% standard errors.}
\label{si-pos-disc}
\end{figure*}

\begin{figure*}[ht!] 
\centering
\includegraphics[trim=0mm 0mm 0mm 0mm, width=\linewidth]{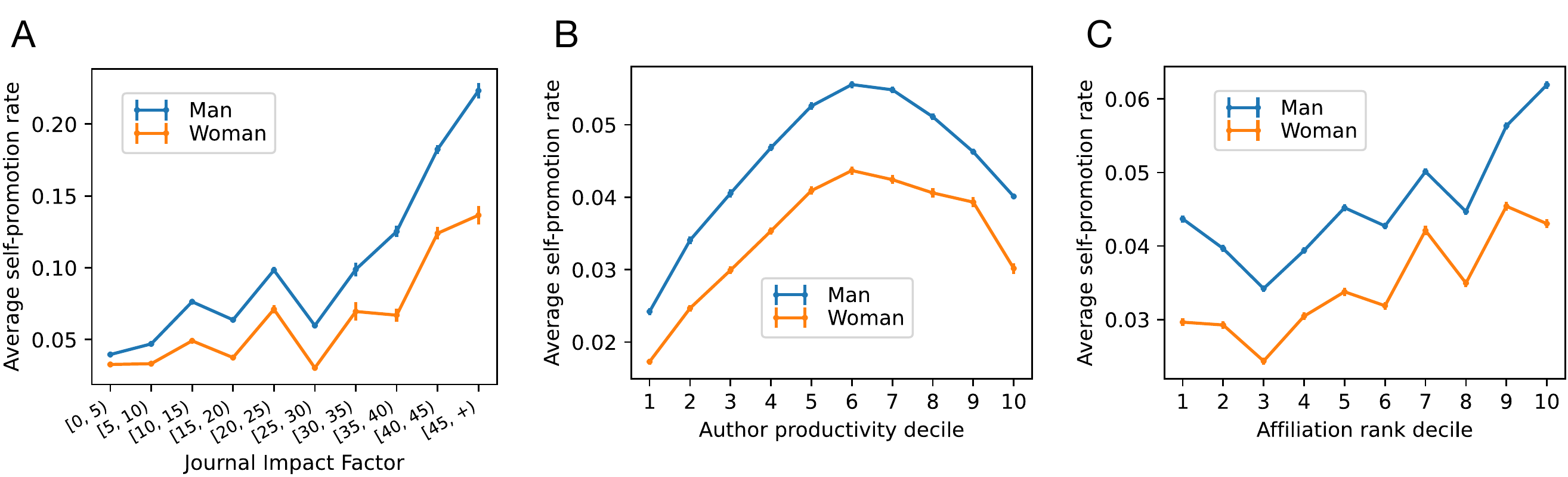}
\caption{The percentage of self-promotion cases by gender in our data as a function of (\textbf{A}) the paper's journal impact factor, (\textbf{B}) the author's productivity decile, measured as the author's previous number of publications, and (\textbf{C}) the author's affiliation rank decile. Confidence intervals represent 95\% standard errors.}
\label{si-fig-gender-pub-raw}
\end{figure*}

\textbf{Effects of control variables: the raw gender difference.}
Fig.~\ref{si-pos-disc}A shows that the average self-promotion probability quadruples from 2013 to 2018 for both genders.
First and last authors are more likely than authors in middle positions to self-promote (Fig.~\ref{si-pos-disc}B), which might be explained by the different roles played by authors in different positions~\cite{vasilevsky2021authorship}. 
Across research fields, self-promotion is much more common among social scientists than among physical, health, and life scientists (Fig.~\ref{si-pos-disc}C).
However, regardless of author roles and research fields, there is a universal gender gap in self-promotion, with men self-advertising between 15\% and 64\% more often than women (Fig.~\ref{si-pos-disc}B-C). 
Fig.~\ref{si-fig-gender-pub-raw} shows that other control variables such as the paper's journal impact factor, the author's affiliation rank, and the author's previous publications all correlate with the chance of self-promotion, regardless of gender. 

\begin{figure*}[ht!] 
\centering
\includegraphics[trim=0mm 0mm 0mm 0mm, width=0.7\linewidth]{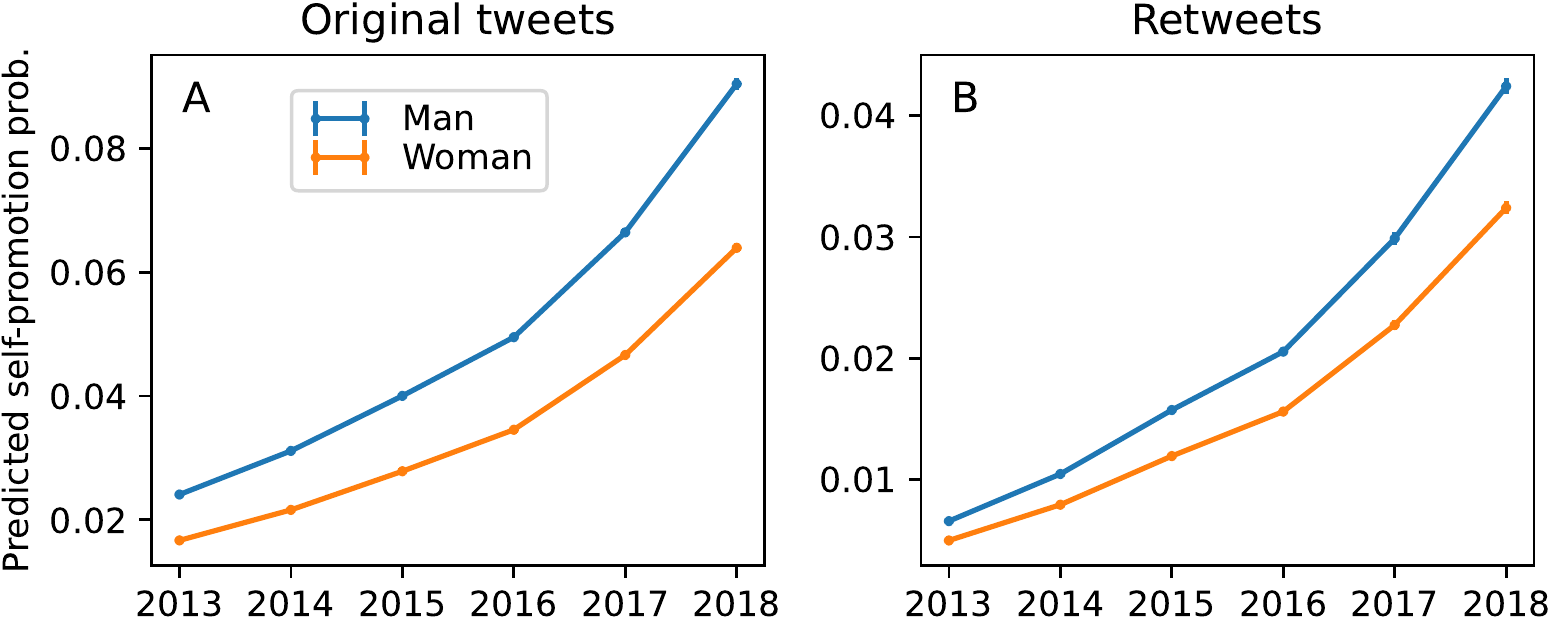}
\caption{The predicted probability of self-promotion after controlling for confounding factors in the mixed-effects logistic regression model as in Fig. 2A in the main text. The two subplots correspond to two types of tweets (original tweets vs. retweets) based on which the binary dependent variable (self-promotion = True) is coded. 
Promotion based on original tweets comes directly from the authors, whereas retweet-based promotion originates from other Twitter users. 
From 2013 to 2018, men self-promote more often than comparable women, across either type of promotional tweets. Error bars indicate 95\% bootstrapped confidence intervals.}
\label{org-vs-retweet-pred}
\end{figure*} 



\clearpage

\section{Supplementary Tables}

\begin{table}[!htbp] \centering
\caption{Results of \textbf{five regression models} predicting whether the author self-promoted the paper. 
Model 1 shows the gender gap without any controls; Model 2 controls for author information; Model 3 adds the characteristics of the paper; Model 4 additionally controls for research topics; and Model 5 further includes paper random effects. The results show that all control variables are correlated with self-promotion and therefore should be included in the final model. Significance levels: *** p$<$0.001, ** p$<$0.01, and * p$<$0.05.}
\label{tab:five-reg}
\resizebox{\textwidth}{!}{
\begin{tabular}{
l @{\extracolsep{10pt}}
S[table-align-text-post=false,table-format=3.6]
S[table-align-text-post=false,table-format=3.6]
S[table-align-text-post=false,table-format=3.6]
S[table-align-text-post=false,table-format=3.6]
S[table-align-text-post=false,table-format=3.6]
}
\\[-1.8ex]\hline 
\hline \\[-1.8ex] 
 & \multicolumn{5}{c}{\textit{Dependent variable:} Self-promotion = True} \\
\cline{2-6} 
\\[-1.8ex] & \textbf{Model 1} & \textbf{Model 2} & \textbf{Model 3} & \textbf{Model 4} & \textbf{Model 5}\\ 
\hline \\[-1.8ex]
  Woman & -0.298$^{***}$ & -0.265$^{***}$ & -0.284$^{***}$ & -0.311$^{***}$ & -0.348$^{***}$ \\ \hline
  Authorship first position &  & 0.337$^{***}$ & 0.315$^{***}$ & 0.339$^{***}$ & 0.361$^{***}$ \\ 
  Authorship middle position &  & -0.579$^{***}$ & -0.649$^{***}$ & -0.584$^{***}$ & -0.638$^{***}$ \\ 
  Authorship solo author &  & 0.894$^{***}$ & 0.938$^{***}$ & 0.686$^{***}$ & 0.828$^{***}$ \\ 
  Affiliation rank &  & -0.052$^{***}$ & -0.048$^{***}$ & -0.053$^{***}$ & -0.048$^{***}$ \\ 
  Affiliation location international &  & 0.046$^{***}$ & 0.080$^{***}$ & 0.080$^{***}$ & 0.052$^{***}$ \\ 
  Author previous num. of publications &  & 0.266$^{***}$ & 0.264$^{***}$ & 0.260$^{***}$ & 0.286$^{***}$ \\ 
  Author previous num. of publications (squared) &  & -0.026$^{***}$ & -0.024$^{***}$ & -0.021$^{***}$ & -0.021$^{***}$ \\ 
  Author log citations &  & 0.020$^{***}$ & -0.002$^{*}$ & -0.013$^{***}$ & -0.031$^{***}$ \\ \hline
  Publication year 2014 &  &  & 0.318$^{***}$ & 0.320$^{***}$ & 0.305$^{***}$ \\ 
  Publication year 2015 &  &  & 0.653$^{***}$ & 0.656$^{***}$ & 0.607$^{***}$ \\ 
  Publication year 2016 &  &  & 0.901$^{***}$ & 0.902$^{***}$ & 0.846$^{***}$ \\ 
  Publication year 2017 &  &  & 1.257$^{***}$ & 1.252$^{***}$ & 1.178$^{***}$ \\ 
  Publication year 2018 &  &  & 1.598$^{***}$ & 1.595$^{***}$ & 1.542$^{***}$ \\
  Number of authors &  &  & -0.001$^{***}$ & -0.001$^{***}$ & -0.002$^{***}$ \\
  Journal impact factor &  &  & 0.034$^{***}$ & 0.035$^{***}$ & 0.033$^{***}$ \\ \hline
  Social Sciences &  &  &  & 0.843$^{***}$ & 0.898$^{***}$ \\ 
  Materials Science &  &  &  & -0.189$^{***}$ & -0.144$^{***}$ \\ 
  Engineering &  &  &  & -0.484$^{***}$ & -0.464$^{***}$ \\ 
  Chemistry &  &  &  & -0.494$^{***}$ & -0.425$^{***}$ \\ 
  Biochemistry Genetics \& Molecular Biology &  &  &  & -0.156$^{***}$ & -0.167$^{***}$ \\ 
  Medicine &  &  &  & -0.317$^{***}$ & -0.279$^{***}$ \\ 
  Nursing &  &  &  & 0.248$^{***}$ & 0.259$^{***}$ \\ 
  Agricultural and Biological Sciences &  &  &  & 0.482$^{***}$ & 0.490$^{***}$ \\ 
  Pharmacology Toxicology \& Pharmaceutics &  &  &  & -0.628$^{***}$ & -0.568$^{***}$ \\ 
  Neuroscience &  &  &  & -0.112$^{***}$ & -0.108$^{***}$ \\ 
  Business  Management and Accounting &  &  &  & -0.200$^{***}$ & -0.181$^{***}$ \\ 
  Economics Econometrics and Finance &  &  &  & -0.476$^{***}$ & -0.490$^{***}$ \\ 
  Chemical Engineering &  &  &  & 0.078$^{***}$ & 0.067$^{***}$ \\ 
  Physics and Astronomy &  &  &  & -0.847$^{***}$ & -0.728$^{***}$ \\ 
  Computer Science &  &  &  & 0.107$^{***}$ & 0.134$^{***}$ \\ 
  Decision Sciences &  &  &  & -0.781$^{***}$ & -0.779$^{***}$ \\ 
  Health Professions &  &  &  & 0.942$^{***}$ & 0.905$^{***}$ \\ 
  Psychology &  &  &  & -0.579$^{***}$ & -0.580$^{***}$ \\ 
  Immunology and Microbiology &  &  &  & -0.170$^{***}$ & -0.175$^{***}$ \\ 
  Dentistry &  &  &  & -1.390$^{***}$ & -1.263$^{***}$ \\ 
  Earth and Planetary Sciences &  &  &  & -0.104$^{***}$ & -0.084$^{***}$ \\ 
  Environmental Science &  &  &  & 0.161$^{***}$ & 0.168$^{***}$ \\ 
  Mathematics &  &  &  & -0.211$^{***}$ & -0.242$^{***}$ \\ 
  Arts and Humanities &  &  &  & -0.226$^{***}$ & -0.229$^{***}$ \\ 
  Energy &  &  &  & -0.441$^{***}$ & -0.369$^{***}$ \\ 
  Veterinary &  &  &  & -0.952$^{***}$ & -0.872$^{***}$ \\ 
  General &  &  &  & 0.376$^{***}$ & 0.388$^{***}$ \\ 
 \hline \\[-1.8ex]
Paper random effects & {No} & {No} & {No} & {No} & {Yes} \\
Intercept & -3.027$^{***}$ & -3.693$^{***}$ & -4.307$^{***}$ & -4.205$^{***}$ & -4.425$^{***}$ \\ \hline
Pseudo-$R^2$ & {0.01} & {0.13} & {0.16} & {0.20} & {0.51} \\
Observations & {11,396,752} & {11,396,752} & {11,396,752} & {11,396,752} & {11,396,752} \\ 
\hline \\[-1.8ex] 
\end{tabular}
}
\end{table}

\begin{table}[!htbp] \centering 
  \caption{Coefficients of all variables in Model 5 (Table~\ref{tab:five-reg}) that predicts \textbf{whether the author is active on Twitter when the paper is published}. The model is fitted to 11,396,752 observations. The negative woman coefficient indicates that women are indeed less likely to be active on Twitter (or have a Twitter account) than comparable men, which explains part of the overall gender gap in self-promotion rates, shown in Table~\ref{tab:five-reg}. Significance levels: *** p$<$0.001, ** p$<$0.01, and * p$<$0.05.}
  \label{si-tab-active-on-tw}
\begin{tabular}{l@{\extracolsep{20pt}}S[table-align-text-post=false,table-format=3.6]} 
\\[-1.8ex]\hline 
\hline \\[-1.8ex] 
 & \multicolumn{1}{c}{\textit{Dependent variable:}} \\ 
\cline{2-2} 
\\[-1.8ex] & \multicolumn{1}{c}{Active on Twitter = True} \\ 
\hline \\[-1.8ex] 
  Woman & -0.134$^{***}$  \\ 
  Authorship first position & 0.295$^{***}$  \\ 
  Authorship middle position & -0.241$^{***}$  \\ 
  Authorship solo author & 0.650$^{***}$  \\ 
  Affiliation rank & -0.071$^{***}$  \\ 
  Affiliation location international & 0.117$^{***}$  \\ 
  Author previous num. of publications & 0.670$^{***}$  \\ 
  Author previous num. of publications (squared) & -0.039$^{***}$  \\ 
  Author log citations & -0.031$^{***}$  \\ 
  Publication year 2014 & 0.498$^{***}$  \\ 
  Publication year 2015 & 0.891$^{***}$  \\ 
  Publication year 2016 & 1.197$^{***}$  \\ 
  Publication year 2017 & 1.511$^{***}$  \\ 
  Publication year 2018 & 1.832$^{***}$  \\ 
  Number of authors & -0.001$^{***}$  \\ 
  Journal impact factor & 0.011$^{***}$  \\ 
  Social Sciences & 0.863$^{***}$  \\ 
  Materials Science & -0.156$^{***}$  \\ 
  Engineering & -0.412$^{***}$  \\ 
  Chemistry & -0.451$^{***}$  \\ 
  Biochemistry Genetics \& Molecular Biology & -0.154$^{***}$  \\ 
  Medicine & -0.084$^{***}$  \\ 
  Nursing & 0.294$^{***}$  \\ 
  Agricultural and Biological Sciences & 0.364$^{***}$  \\ 
  Pharmacology Toxicology \& Pharmaceutics & -0.423$^{***}$  \\ 
  Neuroscience & -0.104$^{***}$  \\ 
  Business  Management and Accounting & -0.135$^{***}$  \\ 
  Economics Econometrics and Finance & -0.470$^{***}$  \\ 
  Chemical Engineering & -0.009 \\ 
  Physics and Astronomy & -0.708$^{***}$  \\ 
  Computer Science & 0.178$^{***}$  \\ 
  Decision Sciences & -0.630$^{***}$  \\ 
  Health Professions & 0.706$^{***}$  \\ 
  Psychology & -0.448$^{***}$  \\ 
  Immunology and Microbiology & -0.197$^{***}$  \\ 
  Dentistry & -1.041$^{***}$  \\ 
  Earth and Planetary Sciences & 0.008 \\ 
  Environmental Science & 0.228$^{***}$  \\ 
  Mathematics & -0.262$^{***}$  \\ 
  Arts and Humanities & -0.251$^{***}$  \\ 
  Energy & -0.369$^{***}$  \\ 
  Veterinary & -0.595$^{***}$  \\ 
  General & 0.100$^{***}$  \\ 
  Intercept & -5.842$^{***}$  \\ \hline
  Pseudo-$R^2$ & 0.42 \\
Observations & {11,396,752} \\ 
\hline 
\hline \\[-1.8ex] 
\end{tabular} 
\end{table} 

\begin{table}[!htbp] \centering 
  \caption{Coefficients of all variables in Model 5 (Table~\ref{tab:five-reg}) fitted to 618,742 observations for which \textbf{the author is active on Twitter at the publication date of the paper}. It shows that the overall gender gap in self-promotion (Table~\ref{tab:five-reg}) is not mainly driven by the gender difference in Twitter presence (Table~\ref{si-tab-active-on-tw}), as the disparity exists even among authors active on Twitter. Note that the quadratic term for \textit{Author previous num. of publications} is dropped because it is linearly correlated with the dependent variable in this subset (including the quadratic term produces similar results). Significance levels: *** p$<$0.001, ** p$<$0.01, and * p$<$0.05.}
  \label{si-tab-rate-on-tw-subset}
\begin{tabular}{l@{\extracolsep{20pt}}S[table-align-text-post=false,table-format=3.6]} 
\\[-1.8ex]\hline 
\hline \\[-1.8ex] 
 & \multicolumn{1}{c}{\textit{Dependent variable:}} \\ 
\cline{2-2} 
\\[-1.8ex] & \multicolumn{1}{c}{Self-promotion = True} \\ 
\hline \\[-1.8ex] 
  Woman & -0.189$^{***}$  \\ 
  Authorship first position & 0.333$^{***}$  \\ 
  Authorship middle position & -0.741$^{***}$  \\ 
  Authorship solo author & 0.375$^{***}$  \\ 
  Affiliation rank & 0.005$^{***}$ \\ 
  Affiliation location international & 0.243$^{***}$  \\ 
  Author previous num. of publications & -0.109$^{***}$  \\
  Author log citations & -0.022$^{***}$  \\ 
  Publication year 2014 & 0.080$^{***}$  \\ 
  Publication year 2015 & 0.171$^{***}$  \\ 
  Publication year 2016 & 0.226$^{***}$  \\ 
  Publication year 2017 & 0.438$^{***}$  \\ 
  Publication year 2018 & 0.650$^{***}$  \\ 
  Number of authors & -0.002$^{***}$  \\ 
  Journal impact factor & 0.022$^{***}$  \\ 
  Social Sciences & 0.322$^{***}$  \\ 
  Materials Science & 0.057$^{*}$ \\ 
  Engineering & -0.099$^{***}$  \\ 
  Chemistry & -0.038 \\ 
  Biochemistry Genetics \& Molecular Biology & -0.095$^{***}$  \\ 
  Medicine & -0.396$^{***}$  \\ 
  Nursing & -0.026 \\ 
  Agricultural and Biological Sciences & 0.371$^{***}$  \\ 
  Pharmacology Toxicology \& Pharmaceutics & -0.268$^{***}$  \\ 
  Neuroscience & -0.030 \\ 
  Business  Management and Accounting & -0.088$^{**}$ \\ 
  Economics Econometrics and Finance & -0.167$^{***}$  \\ 
  Chemical Engineering & 0.122$^{***}$  \\ 
  Physics and Astronomy & -0.235$^{***}$  \\ 
  Computer Science & -0.065$^{**}$ \\ 
  Decision Sciences & -0.312$^{***}$  \\ 
  Health Professions & 0.519$^{***}$  \\ 
  Psychology & -0.349$^{***}$  \\ 
  Immunology and Microbiology & 0.031 \\ 
  Dentistry & -0.698$^{***}$  \\ 
  Earth and Planetary Sciences & -0.093$^{***}$  \\ 
  Environmental Science & 0.122$^{***}$  \\ 
  Mathematics & 0.133$^{***}$  \\ 
  Arts and Humanities & 0.008 \\ 
  Energy & -0.031 \\ 
  Veterinary & -0.616$^{***}$  \\ 
  General & 0.507$^{***}$  \\ 
  Intercept & 0.312$^{***}$  \\ \hline
  Pseudo-$R^2$ & 0.32 \\
Observations & {618,742} \\ 
\hline 
\hline \\[-1.8ex] 
\end{tabular} 
\end{table} 

\begin{table}[!htbp] \centering
  \caption{The same as Model 5 in Table~\ref{tab:five-reg}, but the model is fitted to 12,205,072 observations including \textbf{Woman, Man, and Gender-neutral names}. Authors with gender-neutral names are less likely than those with masculine and feminine names to self-promote their papers. Significance levels: *** p$<$0.001, ** p$<$0.01, and * p$<$0.05.}
  \label{si-tab-rate-unisex}
\begin{tabular}{l@{\extracolsep{20pt}}S[table-align-text-post=false,table-format=3.6]} 
\\[-1.8ex]\hline 
\hline \\[-1.8ex] 
 & \multicolumn{1}{c}{\textit{Dependent variable:}} \\ 
\cline{2-2} 
\\[-1.8ex] & \multicolumn{1}{c}{Self-promotion = True} \\ 
\hline \\[-1.8ex] 
  Woman & -0.348$^{***}$  \\ 
  Unisex & -0.753$^{***}$  \\ 
  Authorship first position & 0.364$^{***}$  \\ 
  Authorship middle position & -0.640$^{***}$  \\ 
  Authorship solo author & 0.844$^{***}$  \\ 
  Affiliation rank & -0.048$^{***}$  \\ 
  Affiliation location international & 0.049$^{***}$  \\ 
  Author previous num. of publications & 0.286$^{***}$  \\ 
  Author previous num. of publications (squared) & -0.021$^{***}$  \\ 
  Author log citations & -0.031$^{***}$  \\   
  Publication year 2014 & 0.305$^{***}$  \\ 
  Publication year 2015 & 0.604$^{***}$  \\ 
  Publication year 2016 & 0.845$^{***}$  \\ 
  Publication year 2017 & 1.176$^{***}$  \\ 
  Publication year 2018 & 1.542$^{***}$  \\ 
  Number of authors & -0.002$^{***}$  \\ 
  Journal impact factor & 0.034$^{***}$  \\ 
  Social Sciences & 0.907$^{***}$  \\ 
  Materials Science & -0.147$^{***}$  \\ 
  Engineering & -0.469$^{***}$  \\ 
  Chemistry & -0.429$^{***}$  \\ 
  Biochemistry Genetics \& Molecular Biology & -0.171$^{***}$  \\ 
  Medicine & -0.274$^{***}$  \\ 
  Nursing & 0.263$^{***}$  \\ 
  Agricultural and Biological Sciences & 0.494$^{***}$  \\ 
  Pharmacology Toxicology \& Pharmaceutics & -0.571$^{***}$  \\ 
  Neuroscience & -0.104$^{***}$  \\ 
  Business  Management and Accounting & -0.178$^{***}$  \\ 
  Economics Econometrics and Finance & -0.493$^{***}$  \\ 
  Chemical Engineering & 0.062$^{***}$  \\ 
  Physics and Astronomy & -0.723$^{***}$  \\ 
  Computer Science & 0.131$^{***}$  \\ 
  Decision Sciences & -0.780$^{***}$  \\ 
  Health Professions & 0.904$^{***}$  \\ 
  Psychology & -0.577$^{***}$  \\ 
  Immunology and Microbiology & -0.177$^{***}$  \\ 
  Dentistry & -1.254$^{***}$  \\ 
  Earth and Planetary Sciences & -0.075$^{***}$  \\ 
  Environmental Science & 0.169$^{***}$  \\ 
  Mathematics & -0.238$^{***}$  \\ 
  Arts and Humanities & -0.228$^{***}$  \\ 
  Energy & -0.374$^{***}$  \\ 
  Veterinary & -0.858$^{***}$  \\ 
  General & 0.392$^{***}$  \\ 
  Intercept & -4.442$^{***}$  \\ \hline
  Pseudo-$R^2$ & 0.52 \\
  Observations & {12,205,072} \\ 
\hline 
\hline \\[-1.8ex] 
\end{tabular} 
\end{table}

\begin{table}[!htbp] \centering 
  \caption{The same as Model 5 in Table~\ref{tab:five-reg} fitted to 9,793,077 observations. The data exclude authors whose names are predicted as \textbf{East Asian ethnicities} based on \textit{Ethnea}~\citep{torvik2016ethnea}. The result shows that our finding is robust for authors whose names display clear gender signals. Significance levels: *** p$<$0.001, ** p$<$0.01, and * p$<$0.05.}
  \label{si-tab-rate-exc-east-asian}
\begin{tabular}{l@{\extracolsep{20pt}}S[table-align-text-post=false,table-format=3.6]} 
\\[-1.8ex]\hline 
\hline \\[-1.8ex] 
 & \multicolumn{1}{c}{\textit{Dependent variable:}} \\ 
\cline{2-2} 
\\[-1.8ex] & \multicolumn{1}{c}{Self-promotion = True} \\ 
\hline \\[-1.8ex] 
 Woman & -0.332$^{***}$  \\ 
  Authorship first position & 0.361$^{***}$  \\ 
  Authorship middle position & -0.623$^{***}$  \\ 
  Authorship solo author & 0.766$^{***}$  \\ 
  Affiliation rank & -0.044$^{***}$  \\ 
  Affiliation location international & 0.096$^{***}$  \\ 
  Author previous num. of publications & 0.280$^{***}$  \\ 
  Author previous num. of publications (squared) & -0.020$^{***}$  \\ 
  Author log citations & -0.038$^{***}$  \\ 
  Publication year 2014 & 0.315$^{***}$  \\ 
  Publication year 2015 & 0.626$^{***}$  \\ 
  Publication year 2016 & 0.868$^{***}$  \\ 
  Publication year 2017 & 1.210$^{***}$  \\ 
  Publication year 2018 & 1.578$^{***}$  \\ 
  Number of authors & -0.002$^{***}$  \\ 
  Journal impact factor & 0.032$^{***}$  \\ 
  Social Sciences & 0.815$^{***}$  \\ 
  Materials Science & -0.078$^{***}$  \\ 
  Engineering & -0.440$^{***}$  \\ 
  Chemistry & -0.415$^{***}$  \\ 
  Biochemistry Genetics \& Molecular Biology & -0.137$^{***}$  \\ 
  Medicine & -0.298$^{***}$  \\ 
  Nursing & 0.228$^{***}$  \\ 
  Agricultural and Biological Sciences & 0.457$^{***}$  \\ 
  Pharmacology Toxicology \& Pharmaceutics & -0.562$^{***}$  \\ 
  Neuroscience & -0.117$^{***}$  \\ 
  Business  Management and Accounting & -0.184$^{***}$  \\ 
  Economics Econometrics and Finance & -0.484$^{***}$  \\ 
  Chemical Engineering & 0.079$^{***}$ \\ 
  Physics and Astronomy & -0.759$^{***}$  \\ 
  Computer Science & 0.144$^{***}$  \\ 
  Decision Sciences & -0.697$^{***}$  \\ 
  Health Professions & 0.870$^{***}$  \\ 
  Psychology & -0.581$^{***}$  \\ 
  Immunology and Microbiology & -0.180$^{***}$  \\ 
  Dentistry & -1.268$^{***}$  \\ 
  Earth and Planetary Sciences & -0.119$^{***}$  \\ 
  Environmental Science & 0.164$^{***}$  \\ 
  Mathematics & -0.292$^{***}$  \\ 
  Arts and Humanities & -0.244$^{***}$  \\ 
  Energy & -0.350$^{***}$  \\ 
  Veterinary & -0.914$^{***}$  \\ 
  General & 0.382$^{***}$  \\ 
  Intercept & -4.299$^{***}$  \\ \hline
  Pseudo-$R^2$ & 0.50 \\
Observations & {9,793,077} \\ 
\hline 
\hline \\[-1.8ex] 
\end{tabular} 
\end{table}

\begin{table}[!htbp] \centering
  \caption{The same as Model 5 in Table~\ref{tab:five-reg}, but the dependent variable is \textbf{whether the author is the first to self-promote the paper}. The model is fitted to 11,396,752 observations. This analysis shows that the gender gap is not driven by potential self-promotion dynamics in mixed-gender teams. Significance levels: *** p$<$0.001, ** p$<$0.01, and * p$<$0.05.}
  \label{si-tab-rate-all-first}
\begin{tabular}{l@{\extracolsep{20pt}}S[table-align-text-post=false,table-format=3.6]} 
\\[-1.8ex]\hline 
\hline \\[-1.8ex] 
 & \multicolumn{1}{c}{\textit{Dependent variable:}} \\ 
\cline{2-2} 
\\[-1.8ex] & \multicolumn{1}{c}{Is the first self-promoter?} \\ 
\hline \\[-1.8ex] 
Woman & -0.323$^{***}$  \\ 
  Authorship first position & 0.352$^{***}$  \\ 
  Authorship middle position & -0.757$^{***}$  \\ 
  Authorship solo author & 0.930$^{***}$  \\ 
  Affiliation rank & -0.049$^{***}$  \\ 
  Affiliation location international & 0.053$^{***}$  \\ 
  Author previous num. of publications & 0.299$^{***}$  \\ 
  Author previous num. of publications (squared) & -0.023$^{***}$  \\ 
  Author log citations & -0.013$^{***}$  \\ 
  Publication year 2014 & 0.287$^{***}$  \\ 
  Publication year 2015 & 0.562$^{***}$  \\ 
  Publication year 2016 & 0.775$^{***}$  \\ 
  Publication year 2017 & 1.051$^{***}$  \\ 
  Publication year 2018 & 1.348$^{***}$  \\ 
  Number of authors & -0.015$^{***}$  \\ 
  Journal impact factor & 0.022$^{***}$  \\ 
  Social Sciences & 0.804$^{***}$  \\ 
  Materials Science & -0.112$^{***}$  \\ 
  Engineering & -0.410$^{***}$  \\ 
  Chemistry & -0.349$^{***}$  \\ 
  Biochemistry Genetics \& Molecular Biology & -0.139$^{***}$  \\ 
  Medicine & -0.253$^{***}$  \\ 
  Nursing & 0.213$^{***}$  \\ 
  Agricultural and Biological Sciences & 0.408$^{***}$  \\ 
  Pharmacology Toxicology \& Pharmaceutics & -0.510$^{***}$  \\ 
  Neuroscience & -0.095$^{***}$  \\ 
  Business  Management and Accounting & -0.140$^{***}$  \\ 
  Economics Econometrics and Finance & -0.440$^{***}$  \\ 
  Chemical Engineering & 0.073$^{***}$  \\ 
  Physics and Astronomy & -0.658$^{***}$  \\ 
  Computer Science & 0.113$^{***}$  \\ 
  Decision Sciences & -0.703$^{***}$  \\ 
  Health Professions & 0.683$^{***}$  \\ 
  Psychology & -0.523$^{***}$  \\ 
  Immunology and Microbiology & -0.142$^{***}$  \\ 
  Dentistry & -1.198$^{***}$  \\ 
  Earth and Planetary Sciences & -0.074$^{***}$  \\ 
  Environmental Science & 0.144$^{***}$  \\ 
  Mathematics & -0.263$^{***}$  \\ 
  Arts and Humanities & -0.207$^{***}$  \\ 
  Energy & -0.279$^{***}$  \\ 
  Veterinary & -0.821$^{***}$  \\ 
  General & 0.268$^{***}$  \\ 
  Intercept & -4.336$^{***}$  \\ \hline
    Pseudo-$R^2$ & 0.86 \\
Observations & {11,396,752} \\ 
\hline 
\hline \\[-1.8ex] 
\end{tabular} 
\end{table}

\begin{table}[!htbp] \centering
  \caption{The same as Model 5 in Table~\ref{tab:five-reg}. The model is fitted to 176,097 observations for which \textbf{the paper is solo-authored}. This analysis shows that our finding is robust for single-authored papers. Significance levels: *** p$<$0.001, ** p$<$0.01, and * p$<$0.05.}
  \label{si-tab-rate-solo-all}
\begin{tabular}{l@{\extracolsep{20pt}}S[table-align-text-post=false,table-format=3.6]} 
\\[-1.8ex]\hline 
\hline \\[-1.8ex] 
 & \multicolumn{1}{c}{\textit{Dependent variable:}} \\ 
\cline{2-2} 
\\[-1.8ex] & \multicolumn{1}{c}{Self-promotion = True} \\ 
\hline \\[-1.8ex] 
  Woman & -0.092$^{***}$  \\ 
  Affiliation rank & -0.025$^{***}$  \\ 
  Affiliation location international & 0.169$^{***}$  \\ 
  Author previous num. of publications & 0.267$^{***}$  \\ 
  Author previous num. of publications (squared) & -0.020$^{***}$  \\ 
  Author log citations & -0.098$^{***}$  \\ 
  Publication year 2014 & 0.417$^{***}$  \\ 
  Publication year 2015 & 0.697$^{***}$  \\ 
  Publication year 2016 & 0.898$^{***}$  \\ 
  Publication year 2017 & 1.214$^{***}$  \\ 
  Publication year 2018 & 1.471$^{***}$  \\ 
  Journal impact factor & 0.025$^{***}$  \\ 
  Social Sciences & 0.497$^{***}$  \\ 
  Materials Science & 0.156 \\ 
  Engineering & -0.119$^{*}$ \\ 
  Chemistry & -0.537$^{***}$  \\ 
  Biochemistry Genetics \& Molecular Biology & -0.067$^{*}$ \\ 
  Medicine & -0.373$^{***}$  \\ 
  Nursing & 0.059 \\ 
  Agricultural and Biological Sciences & 0.237$^{***}$  \\ 
  Pharmacology Toxicology \& Pharmaceutics & -0.315$^{***}$  \\ 
  Neuroscience & 0.036 \\ 
  Business  Management and Accounting & -0.196$^{***}$  \\ 
  Economics Econometrics and Finance & -0.461$^{***}$  \\ 
  Chemical Engineering & 0.227$^{*}$ \\ 
  Physics and Astronomy & -1.124$^{***}$  \\ 
  Computer Science & 0.166$^{***}$  \\ 
  Decision Sciences & -0.418$^{***}$  \\ 
  Health Professions & 0.527$^{***}$  \\ 
  Psychology & -0.414$^{***}$  \\ 
  Immunology and Microbiology & -0.017 \\ 
  Dentistry & -1.490$^{***}$  \\ 
  Earth and Planetary Sciences & -0.152$^{**}$ \\ 
  Environmental Science & 0.091$^{*}$ \\ 
  Mathematics & -1.112$^{***}$  \\ 
  Arts and Humanities & -0.329$^{***}$  \\ 
  Energy & -0.166 \\ 
  Veterinary & -1.433$^{***}$  \\ 
  General & -0.015 \\ 
  Intercept & -2.958$^{***}$  \\ \hline
  Pseudo-$R^2$ & 0.15 \\
Observations & {176,097} \\ 
\hline 
\hline \\[-1.8ex] 
\end{tabular} 
\end{table}

\begin{table}[!htbp] \centering
  \caption{The mean of all control variables by gender in the full data vs. the matched samples in the propensity score matching analysis shown in Table~\ref{si-tab-rate-psm}.}
  \label{si-tab-rate-psm-balance}
\begin{tabular}{|l|rr|rr|}
\hline
\multirow{2}{*}{Feature} & \multicolumn{2}{c|}{\textbf{Before matching}} & \multicolumn{2}{c|}{\textbf{After matching}} \\ \cline{2-5} 
                         & Woman                 & Man                 & Woman                 & Man                \\ \hline
Authorship first position  &    0.19  &    0.15  &    0.19  &    0.19  \\ \hline
Authorship middle position  &    0.66  &    0.64  &    0.66  &    0.66  \\ \hline
Authorship solo author  &    0.01  &    0.02  &    0.01  &    0.01  \\ \hline
Affiliation rank  &    4.35  &    4.28  &    4.35  &    4.38  \\ \hline
Affiliation location international  &    0.65  &    0.65  &    0.65  &    0.65  \\ \hline
Author previous num. of publications  &    4.08  &    5.41  &    4.08  &    4.04  \\ \hline
Author log citations  &    5.93  &    7.84  &    5.93  &    5.85  \\ \hline
Publication year 2014  &    0.15  &    0.15  &    0.15  &    0.15  \\ \hline
Publication year 2015  &    0.17  &    0.17  &    0.17  &    0.17  \\ \hline
Publication year 2016  &    0.18  &    0.18  &    0.18  &    0.18  \\ \hline
Publication year 2017  &    0.19  &    0.18  &    0.19  &    0.19  \\ \hline
Publication year 2018  &    0.18  &    0.19  &    0.18  &    0.18  \\ \hline
Number of authors  &   30.23  &   59.76  &   30.23  &   27.18  \\ \hline
Journal impact factor  &    5.23  &    5.59  &    5.23  &    5.25  \\ \hline
Social Sciences  &    0.09  &    0.06  &    0.09  &    0.08  \\ \hline
Materials Science  &    0.03  &    0.04  &    0.03  &    0.03  \\ \hline
Engineering  &    0.03  &    0.05  &    0.03  &    0.03  \\ \hline
Chemistry  &    0.05  &    0.06  &    0.05  &    0.05  \\ \hline
Biochemistry Genetics \& Molecular Biology  &    0.25  &    0.22  &    0.25  &    0.25  \\ \hline
Medicine  &    0.46  &    0.40  &    0.46  &    0.45  \\ \hline
Nursing  &    0.03  &    0.01  &    0.03  &    0.03  \\ \hline
Agricultural and Biological Sciences  &    0.11  &    0.11  &    0.11  &    0.12  \\ \hline
Pharmacology Toxicology \& Pharmaceutics  &    0.04  &    0.03  &    0.04  &    0.04  \\ \hline
Neuroscience  &    0.05  &    0.05  &    0.05  &    0.05  \\ \hline
Business  Management and Accounting  &    0.01  &    0.01  &    0.01  &    0.01  \\ \hline
Economics Econometrics and Finance  &    0.01  &    0.01  &    0.01  &    0.01  \\ \hline
Chemical Engineering  &    0.02  &    0.02  &    0.02  &    0.02  \\ \hline
Physics and Astronomy  &    0.04  &    0.09  &    0.04  &    0.04  \\ \hline
Computer Science  &    0.01  &    0.03  &    0.01  &    0.02  \\ \hline
Decision Sciences  &    0.00  &    0.00  &    0.00  &    0.00  \\ \hline
Health Professions  &    0.01  &    0.01  &    0.01  &    0.01  \\ \hline
Psychology  &    0.04  &    0.02  &    0.04  &    0.04  \\ \hline
Immunology and Microbiology  &    0.06  &    0.05  &    0.06  &    0.06  \\ \hline
Dentistry  &    0.00  &    0.00  &    0.00  &    0.00  \\ \hline
Earth and Planetary Sciences  &    0.02  &    0.04  &    0.02  &    0.02  \\ \hline
Environmental Science  &    0.04  &    0.05  &    0.04  &    0.05  \\ \hline
Mathematics  &    0.01  &    0.02  &    0.01  &    0.01  \\ \hline
Arts and Humanities  &    0.00  &    0.00  &    0.00  &    0.00  \\ \hline
Energy  &    0.01  &    0.01  &    0.01  &    0.01  \\ \hline
Veterinary  &    0.01  &    0.01  &    0.01  &    0.01  \\ \hline
General  &    0.01  &    0.02  &    0.01  &    0.01  \\ \hline
\end{tabular}
\end{table}

\begin{table}[!htbp] \centering 
  \caption{Logistic regression model that predicts whether the author has self-promoted the paper or not. The model is fitted to 8,051,300 \textbf{matched observations based on the propensity score} using the nearest neighbor matching (one man observation is matched with one woman observation). Significance levels: *** p$<$0.001, ** p$<$0.01, and * p$<$0.05.}
  \label{si-tab-rate-psm}
\begin{tabular}{l@{\extracolsep{20pt}}S[table-align-text-post=false,table-format=3.6]} 
\\[-1.8ex]\hline 
\hline \\[-1.8ex] 
 & \multicolumn{1}{c}{\textit{Dependent variable:}} \\ 
\cline{2-2} 
\\[-1.8ex] & \multicolumn{1}{c}{Self-promotion = True} \\ 
\hline \\[-1.8ex] 
Woman                    &      -0.313$^{***}$    \\
Authorship first position             &      0.338$^{***}$      \\
Authorship middle position           &      -0.619$^{***}$       \\
Authorship solo author               &       0.804$^{***}$      \\
Affiliation rank   & -0.057$^{***}$       \\
Affiliation location international &    0.107$^{***}$     \\
Author previous num. of publications & 0.266$^{***}$        \\
Author previous num. of publications (squared) & -0.020$^{***}$        \\
Author log citations & -0.023$^{***}$        \\
Publication year 2014 & 0.324$^{***}$   \\
Publication year 2015 &  0.657$^{***}$      \\
Publication year 2016 &  0.897$^{***}$      \\
Publication year 2017  & 1.260$^{***}$        \\
Publication year 2018  & 1.603$^{***}$        \\
Number of authors & -0.001$^{***}$       \\
Journal impact factor       &  0.036$^{***}$        \\
Social Sciences & 0.834$^{***}$         \\
Materials Science &  -0.333$^{***}$         \\
Engineering & -0.441$^{**}$       \\
Chemistry & -0.602$^{***}$        \\
Biochemistry Genetics \& Molecular Biology &    -0.224$^{***}$        \\
Medicine                                          &    -0.312$^{***}$        \\
Nursing                                           &   0.209$^{***}$          \\
Agricultural and Biological Sciences           &   0.522$^{***}$          \\
Pharmacology Toxicology \& Pharmaceutics    &   -0.667$^{***}$        \\
Neuroscience                                      &   -0.137$^{***}$         \\
Business  Management and Accounting           &   -0.231$^{***}$         \\
Economics Econometrics and Finance          &    -0.448$^{***}$         \\
Chemical Engineering                            &   0.063$^{***}$        \\
Physics and Astronomy                          &  -0.845$^{***}$        \\
Computer Science                                &   0.118$^{***}$        \\
Decision Sciences                                &  -0.816$^{***}$        \\
Health Professions                               &  0.861$^{***}$  \\
Psychology                                        & -0.582$^{***}$         \\
Immunology and Microbiology                     &  -0.190$^{***}$         \\
Dentistry                                         &  -1.421$^{***}$  \\
Earth and Planetary Sciences                   &  -0.073$^{***}$  \\
Environmental Science                            &  0.148$^{***}$  \\
Mathematics                                       &  -0.267$^{***}$  \\
Arts and Humanities                             &  -0.222$^{***}$  \\
Energy                                            &  -0.414$^{***}$  \\
Veterinary                                        &  -0.851$^{***}$  \\
General                                           &  0.341$^{***}$  \\
Intercept                 &      -4.174$^{***}$    \\
Pseudo-$R^2$ & 0.10 \\
Observations & {8,051,300} \\ 
\hline 
\hline \\[-1.8ex] 
\end{tabular} 
\end{table}


\begin{table}[!htbp]
\centering
\caption{The woman coefficient in a Model 5 (Table~\ref{tab:five-reg}) fitted to all observations of each of the \textbf{four broad disciplines}. Note that a paper can belong to more than one discipline. The result shows that women self-promote less than men across difference research fields. Significance levels: *** p$<$0.001, ** p$<$0.01, and * p$<$0.05.}
\label{si-tab-dis-female}
\begin{tabular}{|c|r|r|}
\hline
Discipline        & \multicolumn{1}{c|}{Number of Observations} & \multicolumn{1}{c|}{Woman Coefficient} \\ \hline
Social Sciences & 0.8M  & -0.195$^{***}$                         \\ \hline
Life Sciences   & 4.1M  & -0.466$^{***}$                         \\ \hline
Physical Sciences & 2.6M & -0.284$^{***}$                        \\ \hline
Health Sciences   & 5.0M  & -0.340$^{***}$                       \\ \hline
\end{tabular}
\end{table}

\begin{table}[!htbp]
\centering
\caption{The same as Table~\ref{si-tab-dis-female}, but each model is fitted to observations where the author is active on Twitter. Significance levels: *** p$<$0.001, ** p$<$0.01, and * p$<$0.05.}
\label{si-tab-dis-female-active}
\begin{tabular}{|c|r|r|}
\hline
Discipline        & \multicolumn{1}{c|}{Number of Observations} & \multicolumn{1}{c|}{Woman Coefficient} \\ \hline
Social Sciences & 77K  & -0.143$^{***}$        \\ \hline
Life Sciences & 193K   & -0.218$^{***}$       \\ \hline
Physical Sciences & 110K & -0.169$^{***}$          \\ \hline
Health Sciences   & 257K  & -0.167$^{***}$         \\ \hline
\end{tabular}
\end{table}


\begin{table}[!htbp] \centering 
  \caption{Negative binomial regression model that predicts \textbf{the total number of tweets of a paper}. The model controls for the author's follower count on Twitter and includes an interaction term between gender and self-promotion. The model is fitted to 618,742 (paper, author) observations where the author is active on Twitter at the publication date of the paper. Note that the quadratic term for \textit{Author previous num. of publications} is dropped because it is linearly correlated with the dependent variable. Paper random effects are not included in the model due to computational infeasibility. Significance levels: *** p$<$0.001, ** p$<$0.01, and * p$<$0.05.}
  \label{si-tab-return-active}
\begin{tabular}{l@{\extracolsep{20pt}}S[table-align-text-post=false,table-format=3.6]} 
\\[-1.8ex]\hline 
\hline \\[-1.8ex] 
 & \multicolumn{1}{c}{\textit{Dependent variable:}} \\ 
\cline{2-2} 
\\[-1.8ex] & \multicolumn{1}{c}{Number of tweets for paper} \\ 
\hline \\[-1.8ex] 
  Woman & 0.148$^{***}$  \\ 
  Self-promotion = True & 1.235$^{***}$  \\ 
  Woman x (Self-promotion = True) & -0.127$^{***}$  \\
  Authorship first position & 0.110$^{***}$  \\ 
  Authorship middle position & 0.290$^{***}$  \\ 
  Authorship solo author & -0.206$^{***}$  \\ 
  Affiliation rank & -0.008$^{***}$  \\ 
  Affiliation location international & 0.069$^{***}$  \\   
  Author previous num. of publications & -0.102$^{***}$  \\ 
  Author log citations & 0.050$^{***}$  \\ 
  Author log follower count & 0.095$^{***}$  \\ 
  Publication year 2014 & 0.239$^{***}$  \\ 
  Publication year 2015 & 0.399$^{***}$  \\ 
  Publication year 2016 & 0.531$^{***}$  \\ 
  Publication year 2017 & 0.797$^{***}$  \\ 
  Publication year 2018 & 0.924$^{***}$  \\ 
  Number of authors & 0.001$^{***}$  \\ 
  Journal impact factor & 0.065$^{***}$  \\ 
  Social Sciences & -0.116$^{***}$  \\ 
  Materials Science & -0.532$^{***}$  \\ 
  Engineering & -0.264$^{***}$  \\ 
  Chemistry & -0.971$^{***}$  \\ 
  Biochemistry Genetics \& Molecular Biology & -0.143$^{***}$  \\ 
  Medicine & -0.035$^{***}$  \\ 
  Nursing & 0.212$^{***}$  \\ 
  Agricultural and Biological Sciences & 0.332$^{***}$  \\ 
  Pharmacology Toxicology \& Pharmaceutics & -0.420$^{***}$  \\ 
  Neuroscience & -0.008 \\ 
  Business  Management and Accounting & -0.408$^{***}$  \\ 
  Economics Econometrics and Finance & -0.129$^{***}$  \\ 
  Chemical Engineering & 0.091$^{***}$  \\ 
  Physics and Astronomy & -0.583$^{***}$  \\ 
  Computer Science & -0.161$^{***}$  \\ 
  Decision Sciences & -0.438$^{***}$  \\ 
  Health Professions & 0.672$^{***}$  \\ 
  Psychology & -0.101$^{***}$  \\ 
  Immunology and Microbiology & -0.007 \\ 
  Dentistry & -0.677$^{***}$  \\ 
  Earth and Planetary Sciences & -0.416$^{***}$  \\ 
  Environmental Science & -0.168$^{***}$  \\ 
  Mathematics & 0.478$^{***}$  \\ 
  Arts and Humanities & -0.151$^{***}$  \\ 
  Energy & -0.588$^{***}$  \\ 
  Veterinary & -0.661$^{***}$  \\ 
  General & 0.449$^{***}$  \\ 
  Intercept & 0.364$^{***}$  \\ 
Observations & {618,742} \\ 
\hline 
\hline \\[-1.8ex] 
\end{tabular} 
\end{table}

\begin{table}[!htbp]
\centering
\caption{The coefficients of Woman and ``Woman x (Self promotion = True)'' in the negative binomial regression model shown in Table~\ref{si-tab-return-active}. Separate models are fitted to data on papers published in each year. Significance levels: *** p$<$0.001, ** p$<$0.01, and * p$<$0.05.}
\label{si-tab-return-active-year}
\begin{tabular}{|c|r|r|}
\hline
 & \multicolumn{2}{c|}{\textit{Dependent variable:} Number of tweets for paper} \\
\cline{2-3}
Publication year   & Woman coeff.    & Woman x (Self promotion = True) coeff. \\ \hline
2013 & 0.272$^{***}$ & -0.170$^{***}$ \\ \hline
2014 & 0.169$^{***}$ & -0.060$^{**\;\,}$ \\ \hline
2015 & 0.158$^{***}$ & -0.131$^{***}$ \\ \hline
2016 & 0.179$^{***}$ & -0.191$^{***}$ \\ \hline
2017 & 0.130$^{***}$ & -0.095$^{***}$ \\ \hline
2018 & 0.110$^{***}$ & -0.066$^{***}$ \\ \hline
\end{tabular}
\end{table}

\begin{table}[!htbp] \centering 
  \caption{The same as Table~\ref{si-tab-return-active}, but is fitted to 20,216 observations for which the paper is solo-authored and the author is active on Twitter. The ``Woman x (Self-promotion = True)'' coefficient is still negative but non-significant, possibly due to a small number of observations. Significance levels: *** p$<$0.001, ** p$<$0.01, and * p$<$0.05.}
  \label{si-return-solo-author} 
\begin{tabular}{l@{\extracolsep{20pt}}S[table-align-text-post=false,table-format=3.6]} 
\\[-1.8ex]\hline 
\hline \\[-1.8ex] 
 & \multicolumn{1}{c}{\textit{Dependent variable:}} \\ 
\cline{2-2} 
\\[-1.8ex] & \multicolumn{1}{c}{Number of tweets for paper} \\ 
\hline \\[-1.8ex] 
  Woman & 0.144$^{***}$  \\ 
  Self-promotion = True & 1.451$^{***}$  \\ 
  Woman x (Self-promotion = True) & -0.053 \\ 
  Affiliation rank & -0.014$^{***}$  \\ 
  Affiliation location international & 0.004 \\ 
  Author previous num. of publications & -0.077$^{***}$  \\ 
  Author log citations & 0.064$^{***}$  \\ 
  Author log follower count & 0.154$^{***}$  \\ 
  Publication year 2014 & 0.211$^{***}$  \\ 
  Publication year 2015 & 0.458$^{***}$  \\ 
  Publication year 2016 & 0.526$^{***}$  \\ 
  Publication year 2017 & 0.790$^{***}$  \\ 
  Publication year 2018 & 0.964$^{***}$  \\ 
  Journal impact factor & 0.035$^{***}$  \\ 
  Social Sciences & -0.039 \\ 
  Materials Science & -0.754$^{***}$  \\ 
  Engineering & -0.273$^{***}$  \\ 
  Chemistry & -0.868$^{***}$  \\ 
  Biochemistry Genetics \& Molecular Biology & 0.072 \\ 
  Medicine & -0.028 \\ 
  Nursing & 0.195$^{**}$ \\ 
  Agricultural and Biological Sciences & 0.250$^{***}$  \\ 
  Pharmacology Toxicology \& Pharmaceutics & 0.188 \\ 
  Neuroscience & 0.311$^{***}$  \\ 
  Business  Management and Accounting & -0.174$^{***}$ \\ 
  Economics Econometrics and Finance & 0.166$^{***}$ \\ 
  Chemical Engineering & 1.121$^{***}$  \\ 
  Physics and Astronomy & -0.103 \\ 
  Computer Science & 0.400$^{***}$  \\ 
  Decision Sciences & -0.353$^{**}$ \\ 
  Health Professions & 1.103$^{***}$  \\ 
  Psychology & 0.139$^{**}$ \\ 
  Immunology and Microbiology & 0.330$^{***}$  \\ 
  Dentistry & -0.473 \\ 
  Earth and Planetary Sciences & -0.070 \\ 
  Environmental Science & -0.031 \\ 
  Mathematics & 0.003 \\ 
  Arts and Humanities & -0.318$^{***}$  \\ 
  Energy & -0.413$^{***}$ \\ 
  Veterinary & -0.337 \\ 
  General & 0.343$^{***}$  \\ 
  Intercept & -0.843$^{***}$  \\ 
Observations & {20,216} \\ 
\hline 
\hline \\[-1.8ex] 
\end{tabular} 
\end{table} 

\begin{table}[!htbp] \centering 
  \caption{The same as Table~\ref{si-tab-return-active}, but the self-promotion variable is defined as an author advertising his or her paper within one day after the paper's publication. Significance levels: *** p$<$0.001, ** p$<$0.01, and * p$<$0.05.}
  \label{si-tab-return-active-diff-def}
\begin{tabular}{l@{\extracolsep{20pt}}S[table-align-text-post=false,table-format=3.6]} 
\\[-1.8ex]\hline 
\hline \\[-1.8ex] 
 & \multicolumn{1}{c}{\textit{Dependent variable:}} \\ 
\cline{2-2} 
\\[-1.8ex] & \multicolumn{1}{c}{Number of tweets for paper} \\ 
\hline \\[-1.8ex] 
  Woman & 0.093$^{***}$  \\ 
  Self-promotion = True & 0.744$^{***}$  \\ 
  Woman x (Self-promotion = True) & -0.089$^{***}$  \\ 
  Authorship first position & 0.180$^{***}$  \\ 
  Authorship middle position & 0.172$^{***}$  \\ 
  Authorship solo author & -0.099$^{***}$  \\ 
  Affiliation rank & -0.007$^{***}$  \\ 
  Affiliation location international & 0.097$^{***}$  \\ 
  Author previous num. of publications & -0.109$^{***}$  \\ 
  Author log citations & 0.039$^{***}$  \\ 
  Author log follower count & 0.135$^{***}$  \\ 
  Publication year 2014 & 0.212$^{***}$  \\ 
  Publication year 2015 & 0.351$^{***}$  \\ 
  Publication year 2016 & 0.487$^{***}$  \\ 
  Publication year 2017 & 0.776$^{***}$  \\ 
  Publication year 2018 & 0.904$^{***}$  \\ 
  Number of authors & 0.001$^{***}$  \\ 
  Journal impact factor & 0.066$^{***}$  \\ 
  Social Sciences & -0.096$^{***}$  \\ 
  Materials Science & -0.490$^{***}$  \\ 
  Engineering & -0.345$^{***}$  \\ 
  Chemistry & -0.990$^{***}$  \\ 
  Biochemistry Genetics \& Molecular Biology & -0.116$^{***}$  \\ 
  Medicine & -0.103$^{***}$  \\ 
  Nursing & 0.155$^{***}$  \\ 
  Agricultural and Biological Sciences & 0.429$^{***}$  \\ 
  Pharmacology Toxicology \& Pharmaceutics & -0.480$^{***}$  \\ 
  Neuroscience & -0.008 \\ 
  Business  Management and Accounting & -0.389$^{***}$  \\ 
  Economics Econometrics and Finance & -0.080$^{***}$  \\ 
  Chemical Engineering & 0.121$^{***}$  \\ 
  Physics and Astronomy & -0.613$^{***}$  \\ 
  Computer Science & -0.147$^{***}$  \\ 
  Decision Sciences & -0.485$^{***}$  \\ 
  Health Professions & 0.726$^{***}$  \\ 
  Psychology & -0.191$^{***}$  \\ 
  Immunology and Microbiology & -0.014 \\ 
  Dentistry & -0.691$^{***}$  \\ 
  Earth and Planetary Sciences & -0.484$^{***}$  \\ 
  Environmental Science & -0.251$^{***}$  \\ 
  Mathematics & 0.557$^{***}$  \\ 
  Arts and Humanities & -0.147$^{***}$  \\ 
  Energy & -0.367$^{***}$  \\ 
  Veterinary & -0.801$^{***}$  \\ 
  General & 0.478$^{***}$  \\ 
  Intercept & 0.599$^{***}$  \\ 
Observations & {618,742} \\ 
\hline 
\hline \\[-1.8ex] 
\end{tabular} 
\end{table} 

\begin{table}[!htbp] \centering 
  \caption{The same as Table~\ref{si-tab-return-active}, but the dependent variable is the number of scientists (``researcher'') who have mentioned each paper. Note that the types of Twitter audiences are categorized by in-house experts from Altmetric.} 
  \label{si-tab-return-active-scientists} 
\begin{tabular}{l@{\extracolsep{20pt}}S[table-align-text-post=false,table-format=3.6]} 
\\[-1.8ex]\hline 
\hline \\[-1.8ex] 
 & \multicolumn{1}{c}{\textit{Dependent variable:}} \\ 
\cline{2-2} 
\\[-1.8ex] & \multicolumn{1}{c}{Number of scientists for paper} \\ 
\hline \\[-1.8ex] 
  Woman & 0.136$^{***}$  \\ 
  Self-promotion = True & 1.272$^{***}$  \\ 
  Woman x (Self-promotion = True) & -0.088$^{***}$  \\ 
  Authorship first position & 0.062$^{***}$  \\ 
  Authorship middle position & 0.304$^{***}$  \\ 
  Authorship solo author & -0.262$^{***}$  \\ 
  Affiliation rank & -0.020$^{***}$  \\ 
  Affiliation location international & 0.045$^{***}$  \\ 
  Author previous num. of publications & -0.189$^{***}$  \\ 
  Author log citations & 0.098$^{***}$  \\ 
  Author log follower count & 0.096$^{***}$  \\ 
  Publication year 2014 & 0.308$^{***}$  \\ 
  Publication year 2015 & 0.456$^{***}$  \\ 
  Publication year 2016 & 0.623$^{***}$  \\ 
  Publication year 2017 & 0.925$^{***}$  \\ 
  Publication year 2018 & 1.099$^{***}$  \\ 
  Number of authors & 0.001$^{***}$  \\ 
  Journal impact factor & 0.062$^{***}$  \\ 
  Social Sciences & -0.091$^{***}$  \\ 
  Materials Science & -0.487$^{***}$  \\ 
  Engineering & -0.453$^{***}$  \\ 
  Chemistry & -0.696$^{***}$  \\ 
  Biochemistry Genetics \& Molecular Biology & 0.177$^{***}$  \\ 
  Medicine & -0.423$^{***}$  \\ 
  Nursing & 0.268$^{***}$  \\ 
  Agricultural and Biological Sciences & 0.401$^{***}$  \\ 
  Pharmacology Toxicology \& Pharmaceutics & -0.568$^{***}$  \\ 
  Neuroscience & -0.036$^{***}$  \\ 
  Business  Management and Accounting & -0.380$^{***}$  \\ 
  Economics Econometrics and Finance & -0.200$^{***}$  \\ 
  Chemical Engineering & 0.122$^{***}$  \\ 
  Physics and Astronomy & -0.254$^{***}$  \\ 
  Computer Science & -0.101$^{***}$  \\ 
  Decision Sciences & -0.149$^{***}$  \\ 
  Health Professions & 0.879$^{***}$  \\ 
  Psychology & -0.113$^{***}$  \\ 
  Immunology and Microbiology & 0.107$^{***}$  \\ 
  Dentistry & -1.440$^{***}$  \\ 
  Earth and Planetary Sciences & -0.385$^{***}$  \\ 
  Environmental Science & -0.130$^{***}$  \\ 
  Mathematics & 0.394$^{***}$  \\ 
  Arts and Humanities & -0.230$^{***}$  \\ 
  Energy & -0.904$^{***}$  \\ 
  Veterinary & -1.053$^{***}$  \\ 
  General & 0.572$^{***}$  \\ 
  Intercept & -1.086$^{***}$  \\ 
Observations & {618,742} \\ 
\hline 
\hline \\[-1.8ex] 
\end{tabular} 
\end{table} 

\begin{table}[!htbp] \centering 
  \caption{The same as Table~\ref{si-tab-return-active}, but the dependent variable is the number of non-scientists (including ``member of the public,'' ``practitioner,'' and ``science communicator'') who have mentioned each paper. Note that the types of Twitter audiences are categorized by in-house experts from Altmetric.}
  \label{si-tab-return-active-non-scientists}
\begin{tabular}{l@{\extracolsep{20pt}}S[table-align-text-post=false,table-format=3.6]} 
\\[-1.8ex]\hline 
\hline \\[-1.8ex] 
 & \multicolumn{1}{c}{\textit{Dependent variable:}} \\ 
\cline{2-2} 
\\[-1.8ex] & \multicolumn{1}{c}{Number of non-scientists for paper} \\ 
\hline \\[-1.8ex] 
  Woman & 0.145$^{***}$  \\ 
  Self-promotion = True & 1.121$^{***}$  \\ 
  Woman x (Self-promotion = True) & -0.125$^{***}$  \\ 
  Authorship first position & 0.126$^{***}$  \\ 
  Authorship middle position & 0.278$^{***}$  \\ 
  Authorship solo author & -0.182$^{***}$  \\ 
  Affiliation rank & -0.010$^{***}$  \\ 
  Affiliation location international & 0.066$^{***}$  \\ 
  Author previous num. of publications & -0.079$^{***}$  \\ 
  Author log citations & 0.040$^{***}$  \\ 
  Author log follower count & 0.092$^{***}$  \\ 
  Publication year 2014 & 0.228$^{***}$  \\ 
  Publication year 2015 & 0.378$^{***}$  \\ 
  Publication year 2016 & 0.495$^{***}$  \\ 
  Publication year 2017 & 0.734$^{***}$  \\ 
  Publication year 2018 & 0.864$^{***}$  \\ 
  Number of authors & 0.001$^{***}$  \\ 
  Journal impact factor & 0.064$^{***}$  \\ 
  Social Sciences & -0.140$^{***}$  \\ 
  Materials Science & -0.621$^{***}$  \\ 
  Engineering & -0.188$^{***}$  \\ 
  Chemistry & -1.176$^{***}$  \\ 
  Biochemistry Genetics \& Molecular Biology & -0.220$^{***}$  \\ 
  Medicine & 0.046$^{***}$  \\ 
  Nursing & 0.293$^{***}$  \\ 
  Agricultural and Biological Sciences & 0.338$^{***}$  \\ 
  Pharmacology Toxicology \& Pharmaceutics & -0.344$^{***}$  \\ 
  Neuroscience & 0.068$^{***}$  \\ 
  Business  Management and Accounting & -0.459$^{***}$  \\ 
  Economics Econometrics and Finance & -0.019 \\ 
  Chemical Engineering & 0.071$^{***}$  \\ 
  Physics and Astronomy & -0.577$^{***}$  \\ 
  Computer Science & -0.213$^{***}$  \\ 
  Decision Sciences & -0.402$^{***}$  \\ 
  Health Professions & 0.646$^{***}$  \\ 
  Psychology & -0.008 \\ 
  Immunology and Microbiology & -0.076$^{***}$  \\ 
  Dentistry & -0.753$^{***}$  \\ 
  Earth and Planetary Sciences & -0.386$^{***}$  \\ 
  Environmental Science & -0.192$^{***}$  \\ 
  Mathematics & 0.285$^{***}$  \\ 
  Arts and Humanities & -0.116$^{***}$  \\ 
  Energy & -0.660$^{***}$  \\ 
  Veterinary & -0.518$^{***}$  \\ 
  General & 0.505$^{***}$  \\ 
  Intercept & 0.036$^{*}$ \\ 
Observations & {618,742} \\ 
\hline 
\hline \\[-1.8ex] 
\end{tabular} 
\end{table} 

\begin{table}[!htbp]
\centering
\begin{tabular}{|c|c|c|c|}
\hline
F1 Score        & \textbf{China} & \textbf{non-China} & \textbf{All Countries} \\ \hline
\textbf{Woman} & 0.61           & 0.90               & 0.83                   \\ \hline
\textbf{Man}   & 0.80           & 0.98               & 0.95                   \\ \hline
\end{tabular}
\caption{The gender prediction accuracy based on authors' self-reported gender in the IOP Publishing data. There are 71,869 authors from China and 361,019 authors from outside of China. Here we list authors from China as a separate group because Chinese names typically do not encode a clear gender signal when written in English characters~\cite{jia2019gender}, and China is the largest single group in this data. As expected, the prediction accuracy for Chinese names is lower than for non-Chinese names, but the overall F1 score is close to 0.9.}
\label{si-tab-iop-gender}
\end{table}

\clearpage

\bibliographystyle{plainnat}
\bibliography{references}